\documentclass[journal]{IEEEtran}

\usepackage{times}
\usepackage{epsfig}
\usepackage{graphicx}
\usepackage{amsmath}
\usepackage{amssymb}
\usepackage{makecell}
\usepackage{colortbl}
\usepackage{multirow}
\usepackage{soul}
\usepackage{adjustbox}
\usepackage{array}
\usepackage{setspace}
\usepackage{lscape}
\usepackage{mathtools}
\usepackage{hyperref}
\usepackage{pgfplots}
\pgfplotsset{compat=newest}

\usepackage{enumitem}
\usepackage{multirow}
\usepackage{rotating}
\usepackage{cite}
\usepackage{amsmath,amssymb,amsfonts}
\usepackage{algorithmic}
\usepackage{graphicx}
\usepackage{textcomp}
\usepackage{caption} 
\usepackage{subcaption}
\usepackage{amssymb}
\usepackage{pifont}
\usepackage{fixfoot,xspace}

\DeclareFixedFootnote*{\dbfootnote}{\url{https://github.com/BiDAlab/ChildCIdb_v1}}
\DeclareFixedFootnote*{\dbfootnotelong}{\url{https://github.com/BiDAlab/ChildCIdbLong}}

\begin{document}

\title{Longitudinal Analysis and Quantitative Assessment of Child Development through Mobile Interaction}

\author{
Juan Carlos Ruiz-Garcia, Ruben Tolosana, Ruben Vera-Rodriguez, Aythami Morales, \\ Julian Fierrez, Javier Ortega-Garcia, Jaime Herreros-Rodriguez\\

\thanks{J.C. Ruiz-Garcia, R. Tolosana, R. Vera-Rodriguez, A. Morales and J. Fierrez are with the Biometrics and Data Pattern Analytics - BiDA Lab, Escuela Politecnica Superior, Universidad Autonoma de Madrid, 28049 Madrid, Spain (e-mail: juanc.ruiz@uam.es; ruben.tolosana@uam.es; ruben.vera@uam.es; aythami.morales@uam.es; julian.fierrez@uam.es; javier.ortega@uam.es).

J. Herreros-Rodriguez is with the Hospital Universitario Infanta Leonor, 28031 Madrid, Spain (e-mail: hrinvest@hotmail.com).}}

\maketitle

\begin{abstract}
    This article provides a comprehensive overview of recent research in the area of Child-Computer Interaction (CCI). The main contributions of the present article are two-fold. First, we present a novel longitudinal CCI database named ChildCIdbLong\dbfootnotelong, which comprises over 600 children aged 18 months to 8 years old, acquired continuously over 4 academic years (2019-2023). As a result, ChildCIdbLong comprises over 12K test acquisitions over a tablet device. Different tests are considered in ChildCIdbLong, requiring different touch and stylus gestures, enabling the evaluation of praxical and cognitive skills such as attentional, visuo-spatial, and executive, among others. In addition to the ChildCIdbLong database, we propose a novel quantitative metric called Test Quality (Q), designed to measure the motor and cognitive development of children through their interaction with a tablet device. In order to provide a better comprehension of the proposed Q metric, popular percentile-based growth representations are introduced for each test, providing a two-dimensional space to compare children's development with respect to the typical age skills of the population.

    The results achieved in the present article highlight the potential of the novel ChildCIdbLong database in conjunction with the proposed Q metric to measure the motor and cognitive development of children as they grow up. The proposed framework could be very useful as an automatic tool to support child experts (e.g., paediatricians, educators, or neurologists) for early detection of potential physical/cognitive impairments during children’s development.
\end{abstract}


\section{Introduction}
\IEEEPARstart{N}{owadays}, the exposure of young children to mobile devices has become nearly universal, with 
a high percentage of children having access to mobile devices before the age of 1 or even having their own device by the age of 4~\cite{Kabali2015}. In particular, screen exposure in children aged 0 to 2 years old has doubled between 1997 and 2014~\cite{Chen2019}, and in children up to 8 years old has increased more than 11 times between 2011 and 2020~\cite{Rideout2020}. This aspect has been exacerbated by the outbreak of COVID-19 in 2020, with studies reporting an alarming increase in the use of digital media by children~\cite{Venigalla2022}. Mobile device use by young children is pervasive and increasing, so the relationship between children's use of mobile devices and their development is only beginning to emerge. For this reason, parents should regulate their children's exposure to mobile devices to enhance their development through educational digital activities and potential bonding through joint use~\cite{Floegel2021}. However, parent-reported duration of mobile device use in young children has been found to have low accuracy, highlighting the need for objective measures in future research~\cite{Radesky2020}. In addition, parents tend to let their children use mobile devices in many situations, such as before going to sleep, after homework, or to keep them calm in public places, among many others~\cite{Kabali2015}.

Despite this massive interaction of children with mobile devices, further research is needed to better understand the impact of mobile device use on young children's learning and development. In this regard, Herodotou analysed in~\cite{Herodotou2018} a total of 19 studies that reported learning and development effects on children aged 2 to 5 years old. Most studies reported positive effects on mathematics, problem-solving, literacy development, and self-learning~\cite{Moosa2020}. However, more longitudinal studies are needed to analyse the evolution of other aspects of children, such as their correct motor and cognitive development~\cite{Danet2022}.

\begin{figure*}[t]
    \begin{center}
       \includegraphics[width=\linewidth]{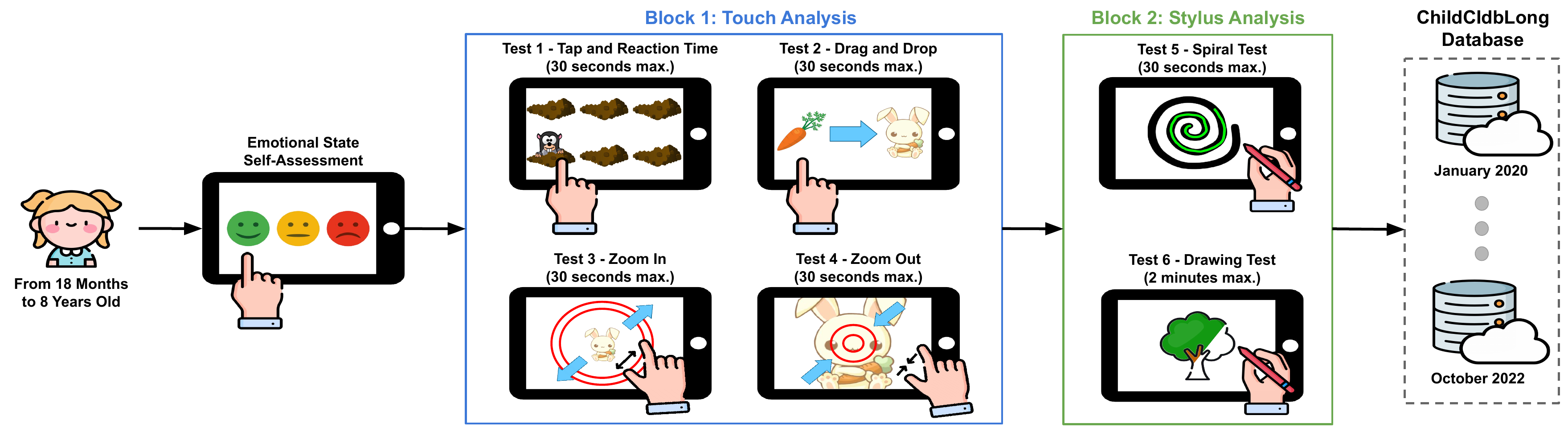}
    \end{center}
    \caption{Graphical representation of the different interfaces designed in ChildCIdbLong, which comprises 6 different data acquisitions from January 2020 to October 2022. Two main acquisition blocks are considered: \textit{i)} touch, and \textit{ii)} stylus.}
    \label{fig:acquisition_process}
\end{figure*}

The present article aims to advance in the Child-Computer Interaction (CCI) research line by proposing a quantitative metric able to automatically measure the motor and cognitive development of children through their interaction with the different tests presented in our unique ChildCIdbLong database. In particular, the main contributions of the present article are:

\begin{itemize}
    \item An in-depth analysis of the state of the art in topics related to: \textit{i)} the interaction of children of different ages with mobile devices; \textit{ii)} the existing longitudinal studies and public databases related to children's exposure to mobile devices and their correct development; \textit{iii)} the variety of gestures that children can perform with mobile devices depending on their age; and \textit{iv)} the most popular tools for measuring the correct motor and cognitive development of children during their growth.

    \item The release of a novel longitudinal database named ChildCIdbLong\dbfootnotelong. As far as we know, this is the largest publicly available longitudinal database to date for research in this area. ChildCIdbLong comprises over 600 children aged 18 months to 8 years old, acquired continuously over 4 academic years (2019/20 to 2022/23). As a result, ChildCIdbLong is composed of over 12K test acquisitions over a tablet device, using both touch and stylus interactions. Fig.~\ref{fig:acquisition_process} provides a graphical representation of the tests considered in ChildCIdbLong.

    \item The proposal of a quantitative metric called Test Quality (Q) to automatically measure the motor and cognitive development of children through their interaction with a tablet device over time. In order to provide a better comprehension of the proposed Q metric, popular percentile-based growth figures are introduced for each test, providing a two-dimensional space to compare children's development with respect to the typical age skills of the population.

    \item A complete experimental analysis of the potential of ChildCIdbLong database to measure the motor and cognitive development of children as they grow up, divided into two approaches: \textit{i)} a general analysis, and \textit{ii)} a longitudinal analysis.
\end{itemize}

The remainder of the article is organised as follows. Sec.~\ref{sec:related_works} summarises an overview of recent studies on children's interaction with mobile devices and their proper development. Sec.~\ref{sec:database} describes all the details of the novel ChildCIdbLong database. Sec.~\ref{sec:methods} describes the proposed Q metric as well as how to calculate it for each ChildCIdbLong test. In Sec.~\ref{sec:analysis}, we compute the potential of the Q metric for measuring the correct motor and cognitive development of children over time. Finally, Sec.~\ref{sec:conclusion} presents the conclusions and future research.

\begin{table*}[t]
    \caption{Comparison of different longitudinal studies focused on the interaction of the children with mobile devices.}
    \label{tab:longitudinal_soa}
    \resizebox{\linewidth}{!}{%
    \begin{tabular}{ccccccc}
    \textbf{Study} & \textbf{Age of Participants} & \textbf{\begin{tabular}[c]{@{}c@{}}\# Participants\\ (Male/Female)\end{tabular}} & \textbf{\begin{tabular}[c]{@{}c@{}}Data Acquisition\\ Periodicity\end{tabular}} & \textbf{\begin{tabular}[c]{@{}c@{}}Acquisition Intervals\\ (Start - End)\end{tabular}} & \textbf{\begin{tabular}[c]{@{}c@{}}Acquired Data\\ (Manual/Automatic)\end{tabular}} & \textbf{Public Database {[}ref{]}} \\ [+10pt] \hline
    \begin{tabular}[c]{@{}c@{}}Brauchli \textit{et al.} (2024)\\ \cite{Brauchli2024}\end{tabular}\rule{0pt}{20pt} & 1-3 Years & \begin{tabular}[c]{@{}c@{}}462\\ (231/231)\end{tabular} & \begin{tabular}[c]{@{}c@{}}4 Time Every\\ Acqusition Interval\end{tabular} & Mar 2021 - Jan 2022 & \begin{tabular}[c]{@{}c@{}}Questionnaries\\ (Manual)\end{tabular} & No \\
    \begin{tabular}[c]{@{}c@{}}McHarg \textit{et al.} (2020)\\ \cite{McHarg2020}\end{tabular}\rule{0pt}{20pt} & 2-3 Years & \begin{tabular}[c]{@{}c@{}}179\\ (100/79)\end{tabular} & \begin{tabular}[c]{@{}c@{}}1 Time Every\\ Acqusition Interval\end{tabular} & \begin{tabular}[c]{@{}c@{}}Children Aged 24 Months\\ Children Aged 36 Months\end{tabular} & \begin{tabular}[c]{@{}c@{}}Questionnaries\\ (Manual)\end{tabular} & Yes~\cite{McHarg2020} \\ 
    \begin{tabular}[c]{@{}c@{}}Poulain \textit{et al.} (2018)\\ \cite{Poulain2018}\end{tabular}\rule{0pt}{20pt} & 2-6 Years & \begin{tabular}[c]{@{}c@{}}527\\ (272/255)\end{tabular} & \begin{tabular}[c]{@{}c@{}}1 Time Every\\ Acqusition Interval\end{tabular} & \begin{tabular}[c]{@{}c@{}}2011 - 2016\\ 2012 - 2017\end{tabular} & \begin{tabular}[c]{@{}c@{}}Questionnaries\\ (Manual)\end{tabular} & Yes~\cite{Poulain2017} \\
    \begin{tabular}[c]{@{}c@{}}Radesky \textit{et al.} (2022)\\ \cite{Radesky2022}\end{tabular}\rule{0pt}{20pt} & 3-5 Years & \begin{tabular}[c]{@{}c@{}}422\\ (224/198)\end{tabular} & \begin{tabular}[c]{@{}c@{}}1 Time\\ Every 3 Months\end{tabular} & Aug 2018 - Jan 2020 & \begin{tabular}[c]{@{}c@{}}Questionnaries\\ (Manual)\end{tabular} & No \\
    \begin{tabular}[c]{@{}c@{}}McNeill \textit{et al.} (2019)\\ \cite{McNeill2019}\end{tabular}\rule{0pt}{20pt} & 3-6 Years & \begin{tabular}[c]{@{}c@{}}185\\ (112/73)\end{tabular} & \begin{tabular}[c]{@{}c@{}}1 Time\\ Every Week\end{tabular} & \begin{tabular}[c]{@{}c@{}}Apr 2015 - Dec 2015\\ Apr 2016 - Dec 2016\end{tabular} & \begin{tabular}[c]{@{}c@{}}Questionnaries\\ (Manual)\end{tabular} & No \\
    \begin{tabular}[c]{@{}c@{}}Konok and Szőke (2022)\\ \cite{Konok2022}\end{tabular}\rule{0pt}{28pt} & 4-9 Years & \begin{tabular}[c]{@{}c@{}}98\\ (54/44)\end{tabular} & \begin{tabular}[c]{@{}c@{}}1 Time Every\\ Acqusition Interval\end{tabular} & \begin{tabular}[c]{@{}c@{}}May 2016 - Aug 2016\\ Dec 2016 - May 2017\\ Nov 2019 - Jan 2020\end{tabular} & \begin{tabular}[c]{@{}c@{}}Questionnaries\\ (Manual)\end{tabular} & Yes~\cite{Konok2022} \\
    \begin{tabular}[c]{@{}c@{}}Byun \textit{et al.} (2013)\\ \cite{Byun2013}\end{tabular}\rule{0pt}{20pt} & 8-11 Years & \begin{tabular}[c]{@{}c@{}}2422\\ (1236/1186)\end{tabular} & \begin{tabular}[c]{@{}c@{}}1 Time Every\\ Acqusition Interval\end{tabular} & \begin{tabular}[c]{@{}c@{}}Oct 2008 - Nov 2008\\ Oct 2010 - Nov 2010\end{tabular} & \begin{tabular}[c]{@{}c@{}}Questionnaries\\ (Manual)\end{tabular} & No \\ [+10pt] \hline
    \textbf{ChildCIdbLong (Present Study)} & \textbf{18 Months - 8 Years} & \textbf{\begin{tabular}[c]{@{}c@{}}615\\ (317/298)\end{tabular}} & \textbf{\begin{tabular}[c]{@{}c@{}}1 Time Every\\ Acqusition Interval\end{tabular}} & \textbf{\begin{tabular}[c]{@{}c@{}}January 2020\\ May 2021\\ October 2021\\ March 2022\\ June 2022\\ October 2022\end{tabular}}\rule{0pt}{44pt} & \textbf{\begin{tabular}[c]{@{}c@{}}Children's Interaction\\ with Mobile Devices\\ (Automatic)\end{tabular}} & \textbf{Yes} \\ [+35pt] \hline
    \end{tabular}
    }
\end{table*}

\section{Related Works}\label{sec:related_works}

\subsection{Child-Computer Interaction (CCI)}\label{subsec:databases}

In recent years, several studies have analysed the interaction of children with different mobile devices and interaction tools (e.g., fingers, keyboard, voice, and pen stylus). In the present article, we focus on studies on works in which children interact using: \textit{i)} their own fingers, and \textit{ii)} a pen stylus. Regarding the studies focused on finger interaction, Crescenzi and Grané presented in~\cite{Crescenzi2019} an analysis of the unstructured interaction of children with a tablet device. In particular, 27 children aged from 14 months to 3 years old interact with 2 apps that allow free drawing and colouring with a small number of interactive screen elements (e.g., colour palette). The results obtained highlight the importance of improving the design of interactive content regarding the mental models and fine motor skills development of children under 3 years old. An interesting article in this research line is the work presented by Kabali \textit{et al.} in~\cite{Kabali2015}. In that work, the authors analysed the use and exposure to mobile devices by 350 children aged 6 months to 4 years. Almost all children (96\%) started using mobile devices before the age of 1, as well as the most popular apps for them are multimedia content applications, such as YouTube and Netflix. For this reason, understanding children's patterns of mobile device usage is crucial to ensure their correct development and behaviour. Similar conclusions have been obtained by Radesky \textit{et al.}~\cite{Radesky2020}. In that work, 346 parents and guardians of children aged 3 to 5 years were recruited to participate in a study for assessing mobile device usage (tablets and smartphones) in children. Statistics show that 35\% of the children had their own device at that age, again mostly using it for multimedia apps (e.g., YouTube, YouTube Kids, and Netflix, among others) and with an average daily usage of about 2 hours per day.

A very interesting article in this research line was presented by Vatavu \textit{et al.}~\cite{Vatavu2015}. The study analysed touch interaction in a dataset with 89 children aged 3 to 6 years old, as well as its relationship to motor skills. The results showed how the touch performance of children improved as they grow, suggesting the designing of touch interfaces adapted to children's limited motor and cognitive skills. The same dataset was considered in other related works~\cite{VeraRodriguez2020, Acien2018}. In~\cite{VeraRodriguez2020}, through an approach based on the lognormality principle, Vera-Rodriguez \textit{et al.} proposed an automatic system able to detect children from adults obtaining over 96\% accuracy. Similar results were obtained by Acien \textit{et al.} in~\cite{Acien2018}, where the authors proposed an active detection approach to classify children from adults based on the global characterisation of touchscreen interaction. Another article contributing to the understanding of children's interaction with touchscreen devices was presented by Nacher \textit{et al.}~\cite{Nacher2018}. The authors assessed the interaction skills of 55 children with Down Syndrome (DS), aged 5-10 years, in the use of multi-touch gestures on touchscreen devices. Despite their limited motor skills, the results showed how DS children were able to perform most of the evaluated multi-touch gestures with high success.

Regarding the use of a pen stylus as an interaction tool with mobile devices, many studies have focused on this research area. Yadav \textit{et al.} analysed the drawing 350 children aged 2-12 years using drawing apps and traditional methods (crayons and watercolours) in~\cite{Yadav2022}. They found that: \textit{i)} children between 2 and 3 years preferred drawing apps with a simpler interface for scribbling and using glowing colours; \textit{ii)} children aged from 4 to 6 years liked to have an eraser feature to correct their drawing; and \textit{iii)} children aged from 7 to 8 years showed significant progress compared to previous age ranges (i.e., use the undo feature, open saved drawings, use different thicknesses, etc.) In~\cite{Mayer2020}, Mayer \textit{et al.} analysed the effects of children's handwriting with paper and pencil, with a stylus on a tablet screen, and typing on a virtual keyboard. A dataset with 145 children aged from 4 to 6 years was captured and divided into three groups: the pencil, stylus, and keyboard groups. The results highlighted that, although the stylus requires higher motor control, the performance of the stylus group did not differ significantly from either the keyboard or the pencil group. A similar article in this research line was presented by Patchan and Puranik~\cite{Patchan2016}. The effectiveness of using tablets as a teaching tool for preschool children (54 participants aged 3 to 5 years) to learn letter-writing skills was investigated by the authors. In particular, the study focused on the impact of extrinsic and intrinsic feedback on children's learning outcomes through 3 ways of interacting: paper and pencil, tablet and finger, or tablet and stylus. The findings indicated that tablet interactions can be beneficial in teaching preschool children to write letters, regardless of the type of feedback provided. Stylus has also been considered by Remi \textit{et al.}~\cite{Remi2015} to analyse the children's motor development through their stylus scribbling skills. For this purpose, the authors considered 60 children aged 3-6 years and the Sigma-Lognormal writing generation model~\cite{VeraRodriguez2020}, concluding that there are significant differences in the model parameters between the ages of children. A different approach was presented by Tabatabaey \textit{et al.} to evaluate children's development through drawing tasks on tablet devices~\cite{Tabatabaey2015}. Drawing data from 631 children aged 6-7 years were analysed to find the relationship between polygonal shape drawing strategies and handwriting performance. The results showed how different drawing strategies influence children's handwriting skills, as well as demonstrated the importance of understanding these relationships for educational purposes. Finally, in order to detect problems in motor skills during children's development, Laniel et al. presented the Pen Stroke Test (PST)~\cite{Laniel2020}, a new measure of fine motor skills able to discriminate between children with attention-deficit/hyperactivity disorder (ADHD) and typically developing children. The study provided preliminary evidence that PST may be useful as a tool for early detection of ADHD.

\subsection{CCI Longitudinal Studies}\label{subsec:longitudinal}

Despite the increased popularity of children interacting with mobile devices, most studies in the literature are based on a single data acquisition in time. As a result, this lack of longitudinal studies does not allow for a proper analysis of the correct children's development during their growth based on the information captured through the interaction with mobile devices~\cite{Herodotou2018, Danet2022}. Table~\ref{tab:longitudinal_soa} shows a comparison of the most relevant longitudinal studies in the literature, ordered by the age of participants, including information such as the number of children considered in each study, the gender balance, and the data acquisition intervals and periodicity, among others. 

In general, most studies focus on whether there is a relationship between children's media exposure and future behavioural difficulties by analysing questionnaires (typically completed by parents/caregivers). For example, McHarg \textit{et al.}~\cite{McHarg2020} analysed the relationship between screen exposure (i.e., television, touchscreens, and computers) and executive function (EF) in children (i.e., EF refers to cognitive processes that involve inhibitory control, working memory, and cognitive flexibility). Parents of 179 children completed similar questionnaires about their children's technology use when they were 2 and 3 years old. Results highlighted how increased screen time in early childhood may have adverse effects on EF development, as well as the need for further research in this area. A similar study was carried out by Radesky \textit{et al.}~\cite{Radesky2022}, exploring the relationship between the use of mobile devices for calming purposes, the EF, and the emotional reactivity of children aged from 3 to 5 years old. In particular, 422 children were assessed by their parents through questionnaires completed once every three months. The findings concluded that there were bidirectional associations, i.e., higher emotional reactivity was associated with increased device use for calming, and higher device use for calming was associated with lower EF scores. An interesting article in this line is the work presented by McNeill \textit{et al.}~\cite{McNeill2019}. This work studied the relationship between the use of electronic applications and media program viewing in preschoolers and their cognitive and psychosocial development. Questionnaire data were collected from 185 preschool children (3-6 years) by their parents, concluding that excessive use of electronic applications can have negative effects on children's cognitive and psychosocial development. Similar conclusions have been obtained in other studies in the literature~\cite{Poulain2018, Konok2022, Byun2013}.

Finally, for completeness, we include in Table~\ref{tab:longitudinal_soa} the description of our novel ChildCIdbLong database, presented in this study. One of the main contributions of this database with respect to other longitudinal studies in the literature is: \textit{i)} we acquire the complete interaction process of the children with the tablet while performing different tests. This allows an in-depth and quantitative analysis of the children, unlike previous approaches in the literature where the only information available is the questionnaires done by parents, which are manual and qualitative; and \textit{ii)} we acquire children aged from 18 months to 8 years old, considering several stages of the children's development. Also, it is important to highlight that the same children are acquired over time (6 acquisition sessions in total), from January 2020 to October 2022, providing a unique CCI database with 615 different children. Sec.~\ref{sec:database} provides more details regarding the ChildCIdbLong database.

\subsection{Children Development and Metrics}\label{subsec:metrics}

In 1973, Jean Piaget defined human development as comprising a sequence of distinct developmental stages, ranging from infancy to adulthood~\cite{Piaget1973}. These developmental stages are based on the evolution of cognitive, physical and social skills. Focusing on physical development, children must acquire rudimentary skills, related to fundamental functions (e.g., walking, running, jumping, etc.), and fine motor skills, related to the execution of complex tasks involving smaller muscle groups, such as manipulating small objects, using scissors or a pencil to write or draw, and performing touch gestures on mobile devices, among many others. 

Children's physical development is a long process in which their motor and cognitive skills constantly improve, enabling them to perform more demanding and complex tasks as they get older. This affects how children interact with mobile devices, as they do not have the same fine motor skills as adults and are not able to perform the same type of interaction. Analysing the finger interaction through touch gestures, there is a wide range of possible gestures that can be performed on screen devices, such as tap, drag, pinch, double tap, and drag-and-drop, among others. In~\cite{Samarakoon2019}, the authors conducted a meta-analysis to analyse previous researches on touch and multi-touch gestures by children aged from 2 to 7 years old. The study provided valuable insights to determine the capabilities and limitations of children in using tactile gestures in the context of CCI. Table~\ref{tab:touch_gestures} provides an overview of the touch gestures performed by children on screen devices, from 2 to 7 years old~\cite{Samarakoon2019}. As can be seen, there is a constant improvement in children's execution of tactile gestures with increasing age. The youngest children (2-3 years) can only perform 4 basic gestures: tap, drag, slide, and pinch (i.e., lower level of cognitive and fine motor skills). Children aged 3-5 years show an improvement compared to 2-3 years but sometimes still need adult supervision when performing some complex gestures. Children older than 5 years can perform all gestures without supervision. Focusing on pen stylus interaction with mobile devices, children's ability to use it depends on some factors, such as age, motor skills and visual perception. Several studies have focused on this research line and have provided information on the developmental milestones associated with children's pen stylus use~\cite{Patchan2016, Cassidy2010, Arif2013}.


\begin{table}[t]
    \caption{Summary of touch gestures performed by children aged from 2 to 7 years old. The gestures highlighted in \textbf{bold} are those considered in the present study. AS refers to ``Adult Supervision''. Table content is adapted from~\cite{Samarakoon2019}.}
    \label{tab:touch_gestures}
    \resizebox{\linewidth}{!}{%
    \begin{tabular}{cccc}
    \multicolumn{4}{c}{\textbf{Age Range (Years)}} \\ \hline
    \multicolumn{1}{c|}{\textbf{2-3}} & \multicolumn{1}{c|}{\textbf{3-4}} & \multicolumn{1}{c|}{\textbf{4-5}} & \textbf{\textgreater{} 5} \\ \hline
    \multicolumn{1}{c|}{\begin{tabular}[c]{@{}c@{}}\textbf{Tap}\\ Drag\\ Slide\\ \textbf{Pinch}\end{tabular}} & \multicolumn{1}{c|}{\begin{tabular}[c]{@{}c@{}}\textbf{Tap}\\ Drag\\ Slide\\ \textbf{Pinch}\\ \textbf{Drag-and-Drop}\\ One Finger Rotation\\ Free Rotate\\ Double Tap (AS)\\ Long Press (AS)\\ Two finger rotation (AS)\end{tabular}} & \multicolumn{1}{c|}{\begin{tabular}[c]{@{}c@{}}Can perform\\ all the gestures,\\ but sometimes\\ may need adult\\ supervision\end{tabular}} & \begin{tabular}[c]{@{}c@{}}Can perform\\ all the gestures\\ without adult \\supervision\end{tabular}
    \end{tabular}
    }
\end{table}

Regarding the procedure followed to measure the correct motor and cognitive development of children as they grow, different tools have been proposed in recent years. Most of them are used by therapists, early intervention specialists, adapted physical education teachers, psychologists, and others who are interested in examining the development of children. One example of these tools is the Bayley-III test kit~\cite{Lawrence2010}. Currently, this third edition is considered the gold standard and is the most widely used tool to identify developmental delays in children aged 1-42 months and to provide information for intervention planning. With this aim, this tool includes 5 different scales related to cognitive, language, motor, social-emotional, and receptive language skills. Applying this tool can take up to 90 minutes. In~\cite{Brown2009}, the authors provided a review of the Movement Assessment Battery for Children-Second Edition (MABC-2). The MABC-2 is a tool, typically comprising 8 different tests for each age range, used to assess and evaluate motor performance in children and adolescents aged from 3 to 16 years old (in 3 age ranges), including manual dexterity, postural balance, and aiming and catching. The study provided valuable insights into the use and effectiveness of the MABC-2 in assessing motor skills disorders and child development. The application time for this tool is around 30 minutes. Another interesting assessment tool for children's gross and fine motor development from birth to 5 years old is the Peabody Developmental Motor Scales (PDMS-3) test kit~\cite{Folio2023}. It contains 6 different tests related to body control and transport, object control, hand manipulation, eye-hand coordination, and physical fitness. Applying this tool takes around 60-90 minutes. Results are provided in 3 composite scores: Gross Motor Index, Fine Motor Index, and Total Motor Index. The reporting system also provides age equivalents, percentile ranks, scaled test scores and composite index scores. As can be observed, most popular tools are time-consuming and the obtained results are highly dependent on the experience and interpretation of the experts using them. For this reason, turns crucial the proposal of automatic and quantitative metrics able to measure the correct motor and cognitive development of children through the use of mobile devices, which is the main purpose of the proposed study. This could be very valuable for therapists and specialists in the area to reduce time and achieve more accurate results.


\begin{table*}[t]
\caption{Relationship between educational levels and age ranges according to the Spanish education system.}
\label{tab:age_range_group}
\resizebox{\linewidth}{!}{%
    \begin{tabular}{c|c|c|c|c|c|c|c}
    \textbf{Educational Level}\rule{0pt}{12pt} & Group 2 & Group 3 & Group 4 & Group 5 & Group 6 & Group 7 & Group 8 \\[+4pt] \hline
    \textbf{Age Range}\rule{0pt}{15pt} & 18 Months - 2 Years & 2-3 Years & 3-4 Years & 4-5 Years & 5-6 Years & 6-7 Years & 7-8 Years \\
    \end{tabular}
}
\end{table*}

\section{ChildCIdbLong Database}\label{sec:database}

\subsection{General Description}\label{subsec:general_description}

As far as we know, ChildCIdbLong is the largest publicly available database to date for research in CCI area. This is an on-going database collected yearly in collaboration with the school GSD Las Suertes in Madrid (Spain). It comprises children aged 18 months to 8 years grouped into 7 different educational levels (Groups 2 to 8) according to the Spanish education system (see details in Table~\ref{tab:age_range_group}). In the proposed framework, children interact with a tablet device using both finger and stylus as an acquisition tool by performing 6 different tests grouped in 2 main blocks: \textit{i)} touch, and \textit{ii)} stylus. Fig.~\ref{fig:acquisition_process} provides a graphical representation of the acquisition process. First of all, the emotional state of the children is captured. In the middle of the screen device, there are 3 faces with different colours (green, yellow, and red) and expressions (happy, normal, and sad). The children must tap one with a finger according to their emotional state. After that, we consider two different test blocks. Each test requires different motor and cognitive skills to be completed correctly within a time range. Next, we briefly describe each of the tests:

\begin{itemize}
    \item \textit{Block 1: Touch Analysis}
    \begin{itemize}
        \item \textbf{Test 1 - Tap and Reaction Time:} there are 6 burrows and 1 mole. Children must tap the mole using only one finger. Then it disappears and reappears in another burrow up to 4 times (30 seconds max). At least 4 taps are needed to complete the test. It requires fine motor skills (tap in a small area) and hand-eye coordination.
        \item \textbf{Test 2 - Drag and Drop:} there is a carrot and a rabbit on the screen. Children must tap the carrot and swipe it to the rabbit using only one finger (30 seconds max). Only 1 tap is needed to complete the test. It combines fine motor skills (tap in a small area), pressure control, hand-eye coordination, and tracking of movement.
        \item \textbf{Test 3 - Zoom In:} there is a small rabbit and 2 circles of different sizes. Children must enlarge the rabbit and put it between circles using 2 fingers (30 seconds max). Only 1 tap/pinch is needed to complete the test. It involves fine motor skills (put the rabbit inside two circles), coordination of the fingers (usually thumb and index finger) for the pinch movement, and accurate perception of the force.
        \item \textbf{Test 4 - Zoom Out:} it is very similar to Test 3, but this time the rabbit must be reduced. 2 fingers and 1 tap/pinch are needed to complete it (30 seconds max).\\\\
    \end{itemize}
    \item \textit{Block 2: Stylus Analysis}
    \begin{itemize}
        \item \textbf{Test 5 - Spiral Test:} using a pen stylus, children must go across the inner part of the black spiral, from the central to the outer part (30 seconds max). Ideally, only 1 stroke is needed to complete the test. It requires precise hand-eye coordination, fine motor skills to control the stylus movement and follow a line without getting off the path, and visual tracking.
        \item \textbf{Test 6 - Drawing Test:} the outline of a tree appears on the screen. Children must colour the whole tree using a pen stylus (2 minutes max). This test involves hand-eye coordination, fine motor skills to control the stylus and stay within the outline of the tree, as well as planning and organisation to colour it properly and fast.
    \end{itemize}
\end{itemize}

All tests were designed considering many of the cognitive and neuromuscular aspects highlighted in the state of the art, e.g., the evolution of children’s gestures with age. In addition, all tests were discussed and approved by neurologists, child psychologists, and educators of the GSD school and their discriminative power was validated in previous experiments~\cite{Tolosana2022-ChildCI, Ruiz-Garcia2024a, Ruiz-Garcia2024b}. For completeness, other children's interesting metadata is also collected such as the previous experience of the children using mobile devices, prematurity (under 37 weeks gestation), cognitive disorders (e.g., developmental delay, ADHD, language disorder, etc.), date of birth, gender, handedness, and academic grades, among others. All this information is always collected with informed parental consent for children's participation in the research project.

\subsection{Longitudinal Data Acquisition}

A preliminary version of ChildCIdbLong database was presented in~\cite{Tolosana2022-ChildCI}, known as ChildCIdb\_v1. This version was collected in January 2020 (just before the COVID-19 outbreak) and comprised a single data acquisition. Since then, 5 more data acquisitions have been collected over time, generating the proposed ChildCIdbLong database\dbfootnotelong. Table~\ref{tab:acquisition_statistics} provides the details of each acquisition session, including the number of children acquired in each educational level as well as the number of tests completed in total (i.e., each child has to perform 6 tests, as described before). In total, ChildCIdbLong comprises over 2.1K children's sessions and over 12.6K children's tests performed and collected in the last 4 academic years (from 2019/20 to 2022/23). It is important to highlight that as ChildCIdbLong is a longitudinal database, the same children were acquired during the 4 academic years in order to study the motor and cognitive evolution of the children over time. As a result, those 2.1K children's sessions were performed by 615 different children, incorporating in every acquisition new children from the youngest groups (Groups 2 to 4) as they were more underrepresented. Also, according to the Spanish education system, an academic year comprises 9 months, from September to June. Between these months, children are assessed at school in 3 main academic evaluations (first, second, and third trimesters). For this reason, our tentative idea was to carry out a new data acquisition every 3 months, thus capturing the evolution of children after each academic evaluation. However, due to the COVID-19 pandemic, this acquisition protocol could not be implemented until the 3rd acquisition (academic year 2021/22).

In Table~\ref{tab:gender_hand_emotion}, we can observe the statistics about gender, handedness, and emotional state associated with each data acquisition. Regarding gender, for all acquisitions, approximately 50\% of the children were male/female. In the case of handedness, around 90\% of the world's population is right-handed~\cite{Papadatou2020}. Left-handed and ambidextrous people are less common, around 10\%~\cite{Denny2007} and 1\%~\cite{Corballis2008} respectively. As can be seen in Table~\ref{tab:gender_hand_emotion}, on average more than 80\% of the children were right-handed, although this is not fully defined until the age of 4-6~\cite{Johnston2009}. In addition, regarding the emotional state of children, at the beginning of the acquisition most children (over 75\%) were in a good mood. However, children under 3 years old may recognise and label some emotions incorrectly~\cite{CrescenziLanna2019}. Therefore, we cannot assume that the youngest children were fully aware of their emotional state when selecting one of the 3 emotional state options.

\subsection{Adults: Control Group}

As we have described in Sec.~\ref{sec:related_works}, children are not able to perform the same type of screen interaction (touch/stylus) as adults due to the different levels of motor and cognitive development. For this reason, and to allow direct comparisons between the motor and cognitive skills of children (in development) and adults (fully developed), a dataset composed of 77 adults was collected as a control group in June 2021, in a single data acquisition. Adults aged from 25 to 65 years old interacted with the tablet device by performing the 6 different tests proposed in ChildCIdbLong. All adults are school workers (e.g. educators, paediatricians, administrators, etc.) without deficits and with normal activity.

\section{Proposed Method: Test Quality (Q)}\label{sec:methods}

As we mentioned in Sec.~\ref{subsec:metrics}, there is a lack of automatic and quantitative metrics that allow to measure the correct motor and cognitive development of children through the use of mobile devices. Current tools are in general time-consuming, manual, and qualitative as it depends on the experience and point of view of the specialist. In order to shed some light on this aspect, in this study we present a new quantitative metric called Test Quality (Q). This metric measures the global quality in which each test is performed, taking into account factors such as the amount of time taken to complete the test and the way of interacting (e.g., number of touches/strokes used, whether the child draws outside the margins or not, etc.). The Q value is a percentage (\%) from 0 to 100, where 100 indicates that the test is performed perfectly and 0 is the opposite. Due to each of the tests designed in ChildCIdbLong requires different motor and cognitive skills and different acquisition tools (touch and stylus), we describe next how to calculate the Q metric for each of the tests presented in ChildCIdbLong. For completeness and reproducibility reasons, the code is also available in GitHub\dbfootnotelong.

\begin{table*}[t]
\caption{Statistics of ChildCIdbLong regarding the number of children that participated in each group and data acquisition.}
\label{tab:acquisition_statistics}
\resizebox{\linewidth}{!}{%
    \begin{tabular}{c|c|c||ccccccc||c|c}
    \textbf{\begin{tabular}[c]{@{}c@{}}ChildCIdb\\ Version\end{tabular}} & \textbf{\begin{tabular}[c]{@{}c@{}}Academic\\ Year\end{tabular}} & \textbf{\begin{tabular}[c]{@{}c@{}}\# Acquisition\\ (Date)\end{tabular}} & \textbf{\begin{tabular}[c]{@{}c@{}}Group 2\\ (18M-2Y)\end{tabular}} & \textbf{\begin{tabular}[c]{@{}c@{}}Group 3\\ (2Y-3Y)\end{tabular}} & \textbf{\begin{tabular}[c]{@{}c@{}}Group 4\\ (3Y-4Y)\end{tabular}} & \textbf{\begin{tabular}[c]{@{}c@{}}Group 5\\ (4Y-5Y)\end{tabular}} & \textbf{\begin{tabular}[c]{@{}c@{}}Group 6\\ (5Y-6Y)\end{tabular}} & \textbf{\begin{tabular}[c]{@{}c@{}}Group 7\\ (6Y-7Y)\end{tabular}} & \textbf{\begin{tabular}[c]{@{}c@{}}Group 8\\ (7Y-8Y)\end{tabular}} & \textbf{\begin{tabular}[c]{@{}c@{}}\# Children's\\ Sessions\end{tabular}} & \textbf{\begin{tabular}[c]{@{}c@{}}\# Children's\\ Tests\end{tabular}} \\ \hline
    \multirow{10}{*}{\rotatebox[origin=c]{90}{\begin{tabular}[c]{@{}c@{}}ChildCIdbLong\\ (Present Study)\end{tabular}}} & 2019/20 & \begin{tabular}[c]{@{}c@{}}1st Acquisition\\ (Jan 2020)\end{tabular} & 18 & 36 & 50 & 66 & 93 & 77 & 98 & 438 & 2,628\\
    & 2020/21 & \begin{tabular}[c]{@{}c@{}}2nd Acquisition\\ (May 2021)\end{tabular} & 40 & 18 & 36 & 51 & 67 & 89 & 75 & 376 & 2,256\\
    & 2021/22 & \begin{tabular}[c]{@{}c@{}}3rd Acquisition\\ (Oct 2021)\end{tabular} & 14 & 45 & 18 & 34 & 49 & 67 & 88 & 315 & 1,890 \\
    & 2021/22 & \begin{tabular}[c]{@{}c@{}}4th Acquisition\\ (Mar 2022)\end{tabular} & 30 & 45 & 17 & 34 & 49 & 66 & 87 & 328 & 1,968\\
    & 2021/22 & \begin{tabular}[c]{@{}c@{}}5th Acquisition\\ (Jun 2022)\end{tabular} & 34 & 44 & 18 & 34 & 48 & 65 & 85 & 328 & 1,968\\
    & 2022/23 & \begin{tabular}[c]{@{}c@{}}6th Acquisition\\ (Oct 2022)\end{tabular} & 5 & 59 & 101 & 18 & 34 & 49 & 65 & 331 & 1,986 \\ \hline
    \multicolumn{1}{c}{} & \multicolumn{1}{c}{} & \multicolumn{1}{c}{} & \multicolumn{1}{c}{} & & & & & & \multicolumn{1}{c||}{\textbf{\# Total}} & 2,116 & 12,696 \rule{0pt}{15pt} 
    \end{tabular}
}
\end{table*}

\begin{table*}[t]
\caption{Statistics of the ChildCIdbLong database regarding the gender, handedness, and emotional state information.}
\label{tab:gender_hand_emotion}
\resizebox{\linewidth}{!}{%
    \begin{tabular}{c|c|cc|cccc|cccc}
    \multirow{3}{*}{\textbf{\begin{tabular}[c]{@{}c@{}}\# Acquisition\\ (Date)\end{tabular}}} & \multirow{3}{*}{\textbf{\begin{tabular}[c]{@{}c@{}}\# Children's\\ Sessions\end{tabular}}}\rule{0pt}{12pt} & \multicolumn{2}{c|}{\textbf{Gender}} & \multicolumn{4}{c|}{\textbf{Handedness}} & \multicolumn{4}{c}{\textbf{Emotional State}} \\[+4pt] \cline{3-12} 
     &  & \textbf{Male}\rule{0pt}{12pt} & \textbf{Female} & \textbf{Right} & \textbf{Left} & \textbf{Both} & \textbf{Unknown} & \textbf{Happy} & \textbf{Normal} & \textbf{Sad} & \textbf{Unknown} \\[+4pt] \hline
    \begin{tabular}[c]{@{}c@{}}1st Acquisition\\ (Jan 2020)\end{tabular}\rule{0pt}{18pt} & 438 & \begin{tabular}[c]{@{}c@{}}219\\ (50\%)\end{tabular} & \begin{tabular}[c]{@{}c@{}}219\\ (50\%)\end{tabular} & \begin{tabular}[c]{@{}c@{}}369\\ (84.2\%)\end{tabular} & \begin{tabular}[c]{@{}c@{}}48\\ (11\%)\end{tabular} & \begin{tabular}[c]{@{}c@{}}17\\ (3.9\%)\end{tabular} & \begin{tabular}[c]{@{}c@{}}4\\ (0.9\%)\end{tabular} & \begin{tabular}[c]{@{}c@{}}342\\ (78.1\%)\end{tabular} & \begin{tabular}[c]{@{}c@{}}13\\ (3\%)\end{tabular} & \begin{tabular}[c]{@{}c@{}}17\\ (3.9\%)\end{tabular} & \begin{tabular}[c]{@{}c@{}}66\\ (15.1\%)\end{tabular} \\[+9pt]
    \begin{tabular}[c]{@{}c@{}}2st Acquisition\\ (May 2021)\end{tabular} & 376 & \begin{tabular}[c]{@{}c@{}}193\\ (51.3\%)\end{tabular} & \begin{tabular}[c]{@{}c@{}}183\\ (48.7\%)\end{tabular} & \begin{tabular}[c]{@{}c@{}}308\\ (81.9\%)\end{tabular} & \begin{tabular}[c]{@{}c@{}}31\\ (8.2\%)\end{tabular} & \begin{tabular}[c]{@{}c@{}}30\\ (8\%)\end{tabular} & \begin{tabular}[c]{@{}c@{}}7\\ (1.9\%)\end{tabular} & \begin{tabular}[c]{@{}c@{}}306\\ (81.4\%)\end{tabular} & \begin{tabular}[c]{@{}c@{}}16\\ (4.3\%)\end{tabular} & \begin{tabular}[c]{@{}c@{}}20\\ (5.3\%)\end{tabular} & \begin{tabular}[c]{@{}c@{}}34\\ (9\%)\end{tabular} \\[+9pt]
    \begin{tabular}[c]{@{}c@{}}3rd Acquisition\\ (Oct 2021)\end{tabular} & 315 & \begin{tabular}[c]{@{}c@{}}165\\ (52.4\%)\end{tabular} & \begin{tabular}[c]{@{}c@{}}150\\ (47.6\%)\end{tabular} & \begin{tabular}[c]{@{}c@{}}272\\ (86.3\%)\end{tabular} & \begin{tabular}[c]{@{}c@{}}29\\ (9.2\%)\end{tabular} & \begin{tabular}[c]{@{}c@{}}10\\ (3.2\%)\end{tabular} & \begin{tabular}[c]{@{}c@{}}4\\ (1.3\%)\end{tabular} & \begin{tabular}[c]{@{}c@{}}239\\ (75.9\%)\end{tabular} & \begin{tabular}[c]{@{}c@{}}21\\ (6.7\%)\end{tabular} & \begin{tabular}[c]{@{}c@{}}16\\ (5.1\%)\end{tabular} & \begin{tabular}[c]{@{}c@{}}39\\ (12.4\%)\end{tabular} \\[+9pt]
    \begin{tabular}[c]{@{}c@{}}4th Acquisition\\ (Mar 2022)\end{tabular} & 328 & \begin{tabular}[c]{@{}c@{}}172\\ (52.4\%)\end{tabular} & \begin{tabular}[c]{@{}c@{}}156\\ (47.6\%)\end{tabular} & \begin{tabular}[c]{@{}c@{}}288\\ (87.8\%)\end{tabular} & \begin{tabular}[c]{@{}c@{}}26\\ (7.9\%)\end{tabular} & \begin{tabular}[c]{@{}c@{}}14\\ (4.3\%)\end{tabular} & \begin{tabular}[c]{@{}c@{}}0\\ (0\%)\end{tabular} & \begin{tabular}[c]{@{}c@{}}246\\ (75\%)\end{tabular} & \begin{tabular}[c]{@{}c@{}}37\\ (11.3\%)\end{tabular} & \begin{tabular}[c]{@{}c@{}}20\\ (6.1\%)\end{tabular} & \begin{tabular}[c]{@{}c@{}}25\\ (7.6\%)\end{tabular} \\[+9pt]
    \begin{tabular}[c]{@{}c@{}}5th Acquisition\\ (Jun 2022)\end{tabular} & 328 & \begin{tabular}[c]{@{}c@{}}175\\ (53.4\%)\end{tabular} & \begin{tabular}[c]{@{}c@{}}153\\ (46.6\%)\end{tabular} & \begin{tabular}[c]{@{}c@{}}289\\ (88.1\%)\end{tabular} & \begin{tabular}[c]{@{}c@{}}24\\ (7.3\%)\end{tabular} & \begin{tabular}[c]{@{}c@{}}13\\ (4\%)\end{tabular} & \begin{tabular}[c]{@{}c@{}}2\\ (0.6\%)\end{tabular} & \begin{tabular}[c]{@{}c@{}}282\\ (85.2\%)\end{tabular} & \begin{tabular}[c]{@{}c@{}}26\\ (7.9\%)\end{tabular} & \begin{tabular}[c]{@{}c@{}}10\\ (3\%)\end{tabular} & \begin{tabular}[c]{@{}c@{}}13\\ (3.9\%)\end{tabular} \\[+9pt]
    \begin{tabular}[c]{@{}c@{}}6th Acquisition\\ (Oct 2022)\end{tabular} & 331 & \begin{tabular}[c]{@{}c@{}}173\\ (52.3\%)\end{tabular} & \begin{tabular}[c]{@{}c@{}}158\\ (47.7\%)\end{tabular} & \begin{tabular}[c]{@{}c@{}}269\\ (81.3\%)\end{tabular} & \begin{tabular}[c]{@{}c@{}}32\\ (9.7\%)\end{tabular} & \begin{tabular}[c]{@{}c@{}}24\\ (7.3\%)\end{tabular} & \begin{tabular}[c]{@{}c@{}}6\\ (1.8\%)\end{tabular} & \begin{tabular}[c]{@{}c@{}}268\\ (81\%)\end{tabular} & \begin{tabular}[c]{@{}c@{}}16\\ (4.8\%)\end{tabular} & \begin{tabular}[c]{@{}c@{}}22\\ (6.6\%)\end{tabular} & \begin{tabular}[c]{@{}c@{}}25\\ (7.6\%)\end{tabular}
    \end{tabular}
}
\end{table*}

\begin{figure}[t]
    \begin{center}
       \includegraphics[width=\linewidth]{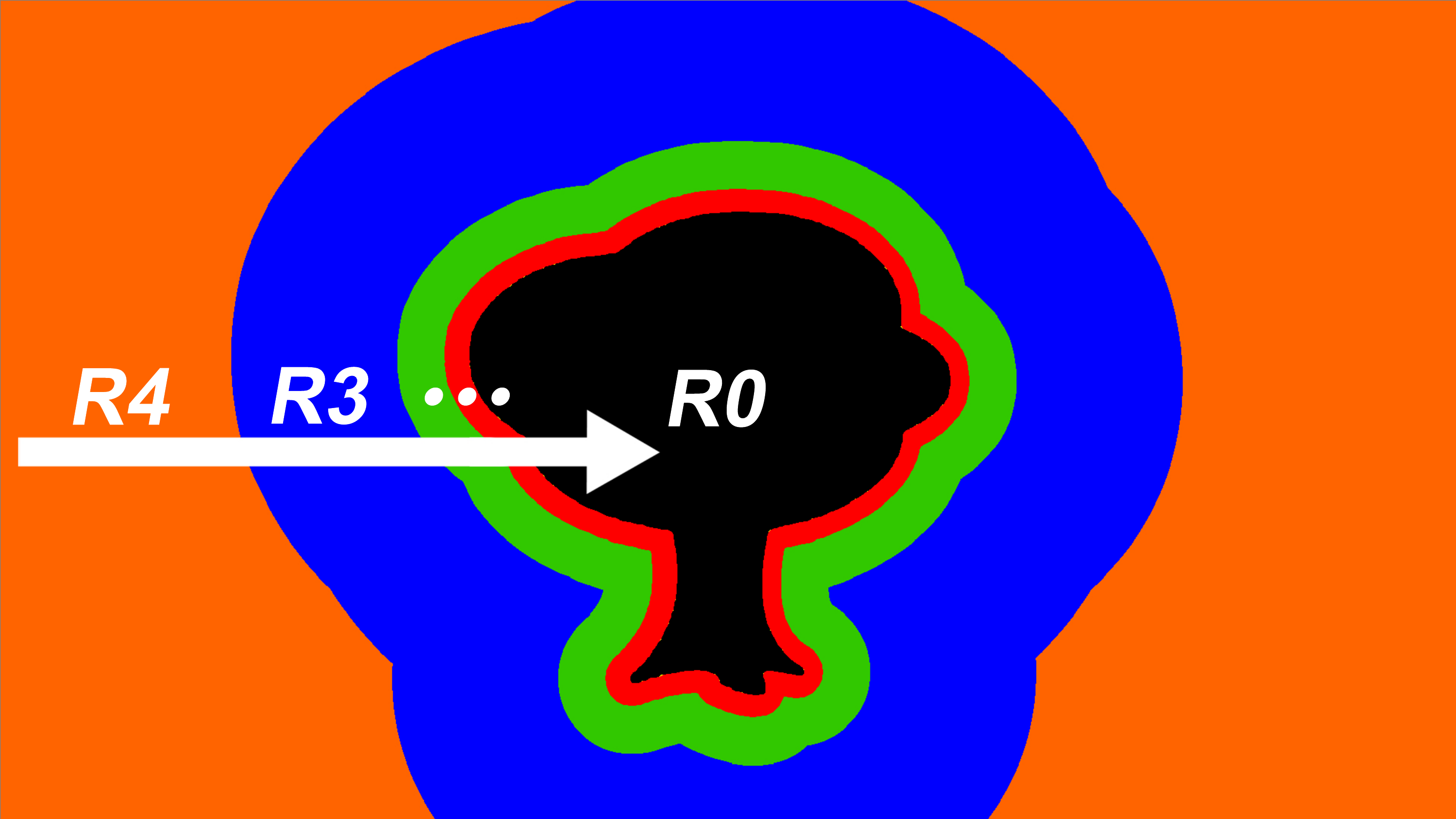}
    \end{center}
    \caption{Regions defined in ``Test 6: Drawing Test'' in order to calculate the proposed Q value. R0, R1, R2, R3, and R4 refer to the different areas of the picture highlighted in black, red, green, blue, and orange regions, respectively.}
    \label{fig:tree_regions}
\end{figure}

\begin{itemize}
    \item \textit{Block 1: Touch Analysis (Tests 1 to 4)}

    The calculation of Q for these tests is indicated by the following equation:

    \begin{equation}\label{eq:q_finger}
        \centering
            Q=\begin{cases}
            \frac{p_{time} + p_{taps}}{2} & \text{, If test is completed}\\ 
            0 & \text{, Otherwise} 
            \end{cases}
    \end{equation}

    The Q value is different to 0 in case the test is completed by the child. In other words, for Test 1 the 4 moles must be touched, for Test 2 the carrot must be moved and touch the rabbit, and for Test 3 and 4 the rabbit must be placed between the two red circles. The amount of time taken to perform the test ($p_{time}$) and the number of finger taps used ($p_{taps}$) are considered in the calculation of the Q value. The longer it takes to perform the test, the lower the Q value will be. The same applies to $p_{taps}$, the Q value will be lower with the number of taps needed to complete the test. As can be seen in the Q value equation, we give the same importance to time ($p_{time}$) and number of taps ($p_{taps}$) in the equation. Table~\ref{tab:q_finger} shows how $p_{time}$ and $p_{taps}$ are calculated for each of the tests (Tests 1 to 4). Both parameters are defined in order to provide percentage values (between 0 and 100). Regarding the definition of the $p_{taps}$ value, for Test 1 this value increases (25) with the number of moles correctly touched ($in_{taps}$) and decreases (5) with the number of taps out of the mole ($out_{taps}$). For Tests 2 to 4, the $p_{taps}$ value decreases with the number of taps performed ($n_{taps}$), as it only requires one tap to complete the test.

    \item \textit{Block 2: Stylus Analysis (Tests 5 to 6)}

    In the following, we describe how to calculate the Q value for each of the tests performed using the stylus:

    \begin{itemize}
        \item \textbf{Test 5: Spiral Test}

        According to the definition of the test, children should go from the inner part of a spiral to the outer part using a pen stylus, always trying to keep it in the black line. However, in our experience, children usually perform the spiral in 4 different ways: from the inner to the outer part, or vice versa, and across the black line or the white line. In all cases, the motor and cognitive skills needed are the same regardless of the way in which the spiral is performed. As a result, we consider 4 spiral templates (one for each case) to calculate the Q value for this test. The 4 spiral templates have been performed by an adult with fully developed motor and cognitive skills by using a single stroke. 
        
        Given a spiral test performed by a child, this is compared with the 4 adult templates using the Dynamic Time Warping (DTW) algorithm~\cite{Fierrez2008, Tolosana2019}. DTW is an algorithm used to measure the similarity between time series, returning a numerical value related to the distance between them. The lower the DTW value is ($d$), the higher the similarity between the time series will be. The following equation indicates the calculation of the Q value using DTW:

        \begin{equation}\label{eq:q_test5}
        \centering
            Q=e^{\frac{-d}{k}} \cdot 100
        \end{equation}

        where $d$ represents the distance between time series using DTW and $k$ refers to the length of the optimal path between the aligned points of the compared time series. This value is finally multiplied by 100 to obtain a Q value in terms of percentage (between 0 and 100).
        
        \item \textbf{Test 6: Drawing Test}

        In the proposed test, children must colour a tree in the best possible way using a pen stylus. We propose the following equation to calculate Q based on 5 different regions of the picture (see Fig.~\ref{fig:tree_regions}):

        \begin{equation}\label{eq:q_test6}
            \centering
            Q = R0 - (R1 + R2 + R3 + R4)
        \end{equation}

        where Rx is related to the area painted in that particular region of the picture. R0 (black) represents the inner region of the tree, i.e., all pixels within the tree outline. It is a percentage value between 0-100\%, where 100\% means all pixels of R0 are coloured (fully coloured tree) and 0\% means none (uncoloured tree). R1 (red), R2 (green), R3 (blue), and R4 (orange) represent the outer regions of the tree. If children colour in any of these regions, a penalty is applied. In particular, the penalty value is higher according to the distance with respect to the R0 region: R4 up to 40\%, R3 up to 30\%, R2 up to 20\%, and R1 up to 10\%. If the Q value obtained is negative (i.e., more area is coloured outside the tree than inside), the value is set to 0\%.
    \end{itemize}
\end{itemize}

\begin{figure}[t]
    \begin{center}
       \includegraphics[width=\linewidth]{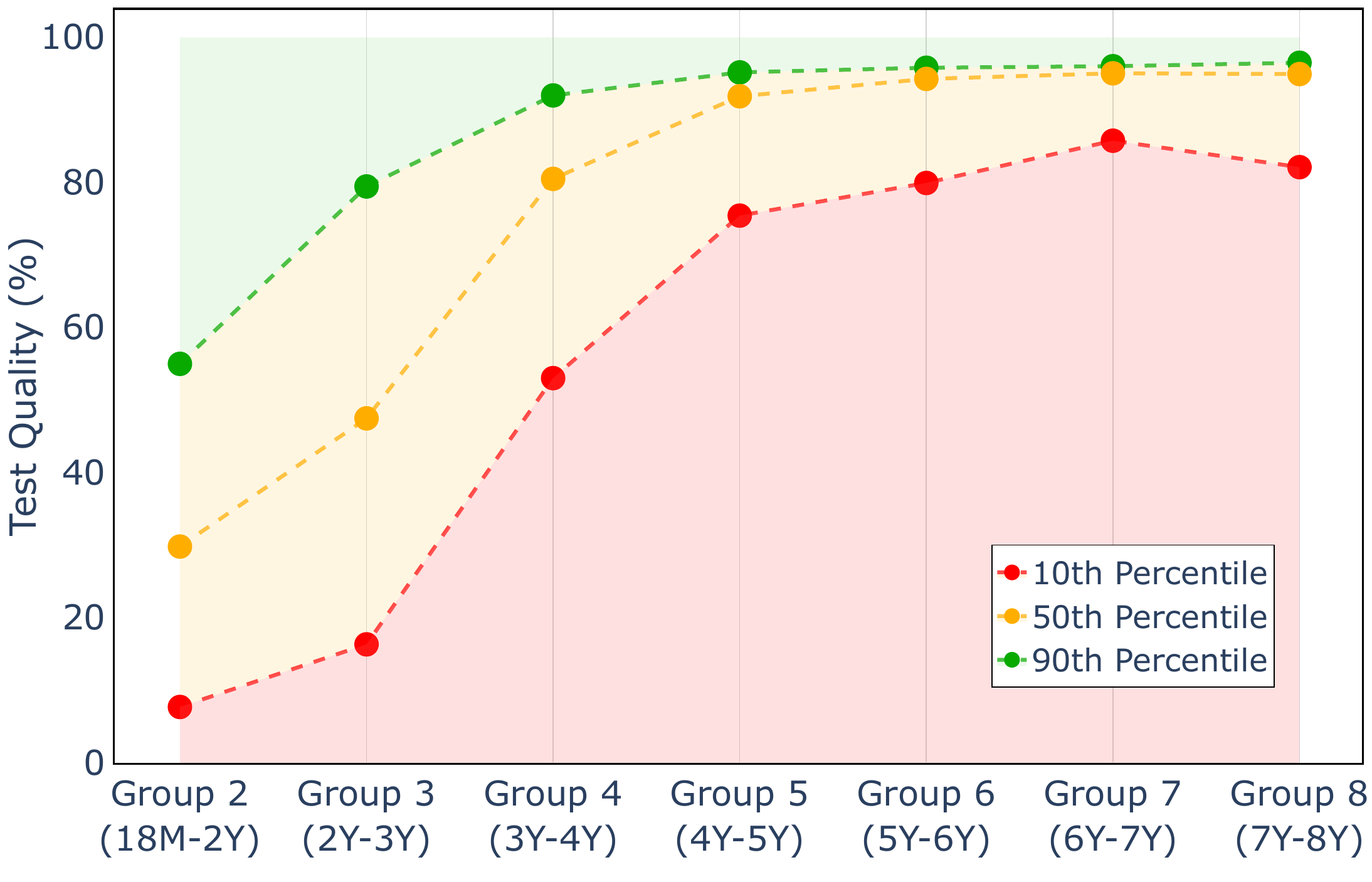}
    \end{center}
    \caption{Percentile-based growth representation in terms of the proposed Q metric for ``Test 6: Drawing Test'' of the ChildCIdbLong.}
    \label{fig:percentiles}
\end{figure}

In addition to the proposed Q metric designed for each test of ChildCIdbLong, we also consider percentile-based growth representations that can help therapists and specialists in the field to assess whether a child's motor and cognitive development is correct or not along time. According to the World Health Organization (WHO), it's critical the use of 3rd, 10th, 50th, 90th, and 97th percentiles to define optimal growth values (e.g., height-, length-, weight-, and body mass index-for-age) as standards based on worldwide data of healthy children~\cite{Onis2007}. For this reason, in the proposed study we calculate the 15th, 50th, and 90th percentiles of the proposed Q metric to generate a growth representation for each ChildCIdbLong test. The 3rd and 97th percentiles are not considered so far as we do not have enough data to give a robust growth chart at these limits. The proposal of the Q value for each test, together with the proposed graphical representation in terms of percentile, can be very beneficial to quantitatively measure the motor and cognitive development of children through the use of mobile devices. An example of this can be observed in Fig.~\ref{fig:percentiles}, where we represent the proposed Q value in terms of the percentile-based growth for Test 6 of ChildCIdbLong. For example, for the Group 3 (2Y-3Y), the 10th Percentile curve means that the 10\% of the child population at that age (2-3 years) achieve a Q value below 17\% whereas the remainder 90\% of the child population obtains a higher Q value.


\section{Experimental Framework}\label{sec:analysis}

This section analyses the potential of the proposed Q metric using the novel ChildCIdbLong database. In particular, we focus on two different analyses: \textit{i)} a general analysis of the Q-values distributions for each of the 6 tests and educational levels (Group 2 to 8) considered in ChildCIdbLong, and \textit{ii)} a longitudinal analysis of the Q values while the children grow up, analysing their motor and cognitive development over time using the samples collected at each data acquisition for the last 4 academic years.

\begin{table}[t]
    \caption{Equations defined to calculate $p_{time}$ and $p_{taps}$ for Tests 1 to 4. $t_{max}$: maximum test time; $t_{real}$: time taken to perform the test; $in_{taps}$ and $out_{taps}$: number of taps on and off the mole, respectively; $n_{taps}$: total number of finger taps used.}
    \label{tab:q_finger}
    \adjustbox{width=0.5\textwidth}{
        \begin{tabular}{c|c|c}
         & $\pmb{p_{time}}$ & $\pmb{p_{taps}}$ \\ [+3pt] \hline
        \textit{\begin{tabular}[c]{@{}c@{}}Test 1: Tap and\\ Reaction Time\end{tabular}} & \multirow{4}{*}{$\frac{t_{max} - t_{real}}{t_{max}} \cdot 100$} & $in_{taps} \cdot 25 - out_{taps} \cdot 5$ \\ [+5pt] \cline{1-1} \cline{3-3}
        \textit{\begin{tabular}[c]{@{}c@{}}Test 2: Drag\\ and Drop\end{tabular}} &  & \multirow{3}{*}{$ 
        \begin{cases}
            0 & \text{, } n_{taps}=0 \\ 
            \frac{100}{n_{taps}} & \text{, } n_{taps} > 0 
        \end{cases}$} \\ [+3pt] \cline{1-1}
        \textit{Test 3: Zoom In} &  &  \\ \cline{1-1}
        \textit{Test 4: Zoom Out} &  & \\ \hline
        \end{tabular}
    }
\end{table}

\subsection{General Analysis}
\begin{figure*}[t]
    \begin{center}
       \includegraphics[width=\linewidth]{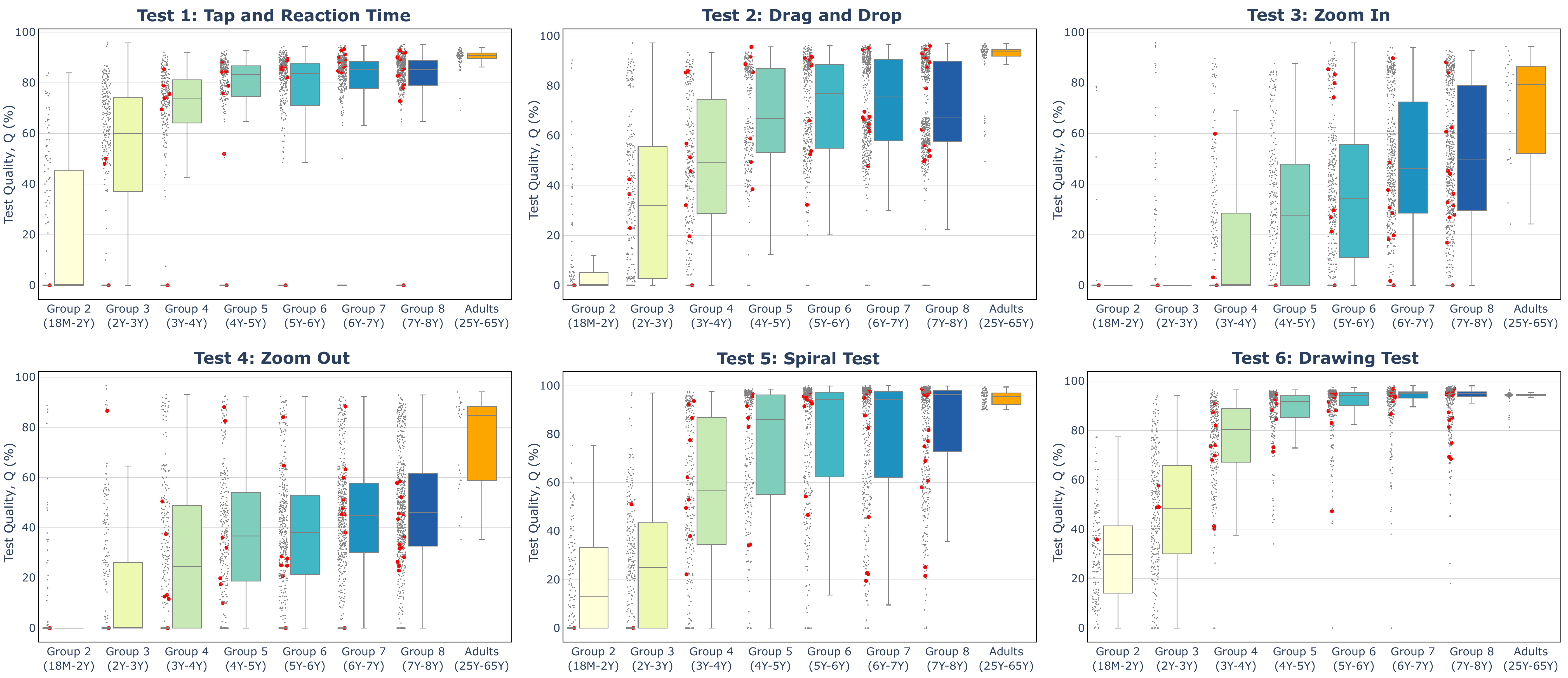}
    \end{center}
    \caption{Graphical representation of the Q values achieved in each of the tests and age groups considered in ChildCIdbLong. Red points refer to children with Non-Typical Development (NTD). The inner horizontal line of the box represents the median value. The lower and upper ends of the box represent the Q1 and Q3 quartiles, respectively. Whiskers represent the outliers.}
    \label{fig:general}
\end{figure*}

Fig.~\ref{fig:general} provides a graphical representation of the Q values achieved in each of the tests and age groups considered in ChildCIdbLong. Each box plot contains boxes, whiskers, and points. The inner horizontal line of the box represents the median value, and the lower and upper ends represent the Q1 and Q3 quartiles, respectively. The whiskers refer to the outliers (i.e., values over the Q3 quartile and under the Q1 quartile). Each point refers to one sample (i.e., a test performed by a child). Black points represent children with Typical Development (TD) whereas red points represent children with Non-Typical Development (NTD), i.e., children without/with developmental disorders such as developmental delay, ADHD, language disorder, etc.

As we mentioned in Sec.~\ref{subsec:general_description}, each test proposed in ChildCIdbLong requires different gestures (tap, drag-and-drop, pinch, line-following, and drawing) and different motor and cognitive skills to be completed correctly. This aspect can be observed in Fig.~\ref{fig:general}, as the proposed Q value increases in all tests as children get older and develop their motor and cognitive skills. In addition, for completeness, we include in Table~\ref{tab:average_general} the average results of the proposed Q metric for each test and age group of ChildCIdbLong, including the control group (Adults). In general, these results prove the correct definition of the tests considered in ChildCIdbLong as well as the proposed Q metric to evaluate the motor and cognitive skills of the children as they grow up.

\begin{table*}[t]
    \caption{Average Q value achieved in each test and age group of the ChildCidbLong database, including the control group (Adults). For completeness, we also specify the hand or stylus gesture required to complete each test, as described in Sec.~\ref{subsec:general_description}. The Q values of Groups 6 to 8 are shown together as similar average results are achieved between them.}
    \label{tab:average_general}
    \centering
    \resizebox{\textwidth}{!}{ 
        \begin{tabular}{c|cccc|cc}
         & \multicolumn{4}{c|}{\textbf{Block 1: Touch Analysis}} & \multicolumn{2}{c}{\textbf{Block 2: Stylus Analysis}} \\ \cline{2-7} 
         & \textbf{\begin{tabular}[c]{@{}c@{}}Test 1: Tap and Reaction Time\\ (Tap)\end{tabular}} & \textbf{\begin{tabular}[c]{@{}c@{}}Test 2: Drag and Drop\\ (Drag-and-Drop)\end{tabular}} & \textbf{\begin{tabular}[c]{@{}c@{}}Test 3: Zoom-In\\ (Pinch)\end{tabular}} & \textbf{\begin{tabular}[c]{@{}c@{}}Test 4: Zoom-Out\\ (Pinch)\end{tabular}} & \textbf{\begin{tabular}[c]{@{}c@{}}Test 5: Spiral Test\\ (Line-Following)\end{tabular}} & \textbf{\begin{tabular}[c]{@{}c@{}}Test 6: Drawing Test\\ (Drawing)\end{tabular}} \\ \hline
        \textbf{\begin{tabular}[c]{@{}c@{}}Group 2\\ (18M-2Y)\end{tabular}} & 20.5 \% & 8.6 \% & 2.2 \% & 6.4 \% & 18.9 \% & 29.8 \% \\ \hline
        \textbf{\begin{tabular}[c]{@{}c@{}}Group 3\\ (2Y-3Y)\end{tabular}} & 50.5 \% & 34.2 \% & 5.9 \% & 14.6 \% & 26.4 \% & 47.5 \% \\ \hline
        \textbf{\begin{tabular}[c]{@{}c@{}}Group 4\\ (3Y-4Y)\end{tabular}} & 65.8 \% & 49.8 \% & 16.9 \% & 28.4 \% & 56.1 \% & 75.9 \% \\ \hline
        \textbf{\begin{tabular}[c]{@{}c@{}}Group 5\\ (4Y-5Y)\end{tabular}} & 70.8 \% & 68.3 \% & 29.9 \% & 37.7 \% & 73.9 \% & 87.3 \% \\ \hline
        \textbf{\begin{tabular}[c]{@{}c@{}}Groups 6 to 8\\ (5Y-8Y)\end{tabular}} & 70.1 \% & 72.1 \% & 44.9 \% & 43.6 \% & 80.6 \% & 91.2 \% \\ \hline
        \textbf{\begin{tabular}[c]{@{}c@{}}Adults\\ (25Y-65Y)\end{tabular}} & 89.8 \% & 89.7 \% & 71.0 \% & 73.9 \% & 94.7 \% & 95.6 \%
        \end{tabular}
    }
\end{table*}

Before analysing the results obtained by the different groups of children, it is important to note that the Q metric, as described in Sec.~\ref{sec:methods}, takes into account 2 factors: \textit{i)} the amount of time taken to complete the test, and \textit{ii)} the way children interact with the test (number of taps/strokes, the child colours inside the spiral/tree or not, etc.). This means that children may not be able to complete the test in the required time due to the lack of comprehension to do it properly or just because their cognitive and motor skills are not developed yet. From the results observed in Fig.~\ref{fig:general} and Table~\ref{tab:average_general}, it is evident that most children between 18 months and 2 years old (Group 2) are not able to complete any tests correctly (i.e., average Q values below 20\% and 30\% for the touch and stylus tests, respectively). We hypothesise that children at this age are still in the early stages of both cognitive and motor development. Nevertheless, they seem to be able to perform somehow the tap gesture using the finger (Test 1) and drawing (or at least scrawling) with the stylus (Test 6), being the tests that obtain the best Q values, although far from the control group (i.e., 20-30\% vs. 90-95\% Q values). Regarding children aged 2 to 3 years old (Group 3), we can observe some improvements compared to children in Group 2 when performing the drag-and-drop gesture (Test 2) with the finger, as well as improving their performance in the tap gesture (Test 1) or drawing with the stylus (Test 6). However, they are not able to perform correctly more complex fine motor gestures yet, such as the pinch with two fingers (Tests 3 and 4) or following a line with the stylus without getting off the path (Test 5), achieving Q values lower than 30\% on average. 

\begin{figure*}[t]
    \begin{center}
       \includegraphics[width=\linewidth]{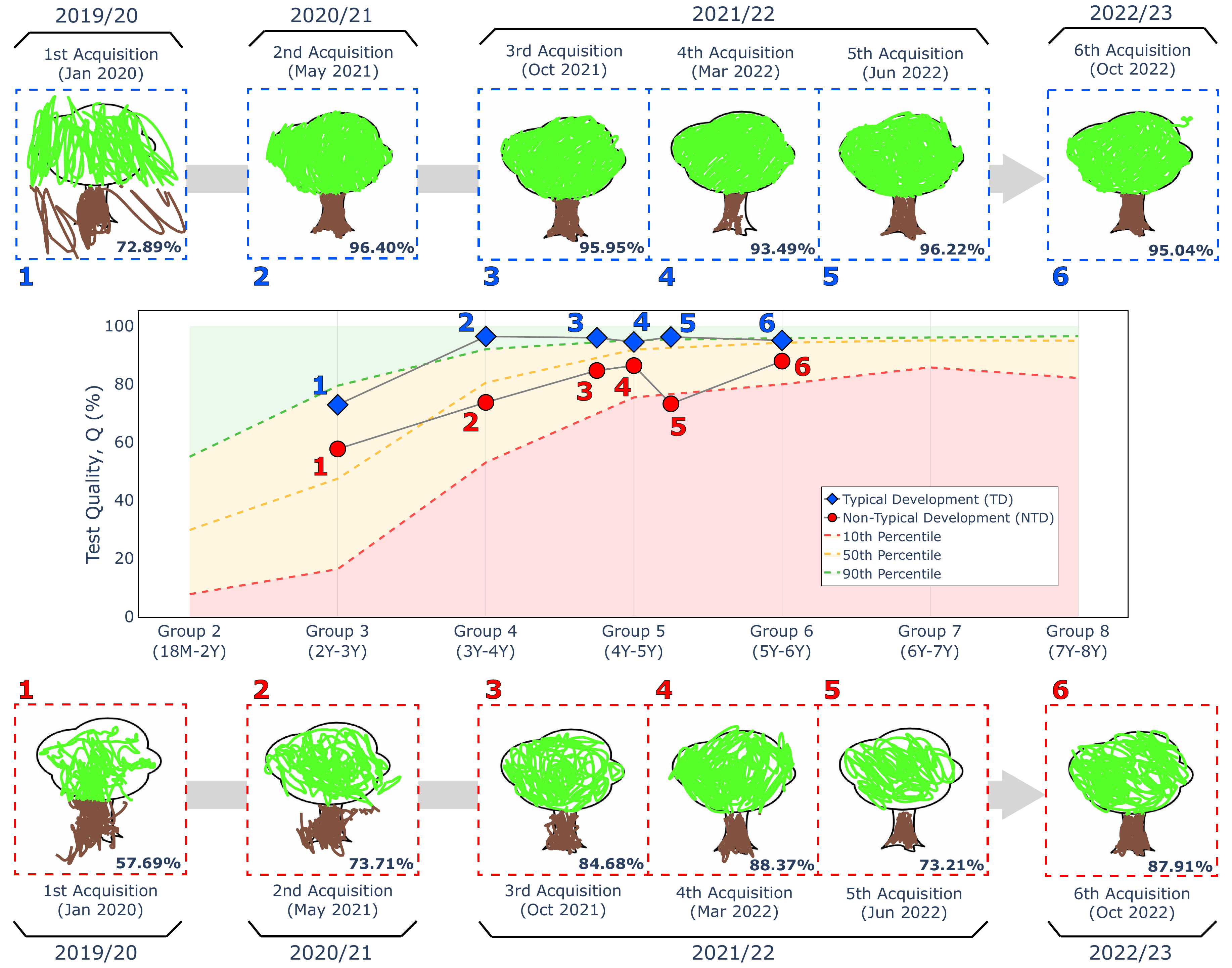}
    \end{center}
    \caption{Examples of the evolution of the proposed Q values achieved in Test 6 for two different children: \textit{i)} a TD child without apparent physical/cognitive impairment, and \textit{ii)} a NTD child with special educational needs and an expressive language disorder. We also include on top/bottom of the figure the 6 graphical tree's executions, each corresponding with one acquisition in time.}
    \label{fig:longitudinal}
\end{figure*}

Analysing children aged 3 to 5 years old (Groups 4 and 5), there is a considerable improvement in both touch and stylus gestures. Focusing on stylus gestures, children begin to develop more precise fine motor skills around this age, which allow them to perform more accurate strokes and even begin to write~\cite{Puranik2011}. This fact is demonstrated in Fig.~\ref{fig:general} and Table~\ref{tab:average_general}, as the Q values for Test 5 (line-following gesture) and Test 6 (drawing gesture) improve significantly with respect to previous age groups, reaching average Q values of 73.9\% and 87.63\%, respectively. These results indicate that at this age the motor and cognitive skills of the children are more developed, being able to follow the spiral line and colour the tree in less time and without going out of outline as often. Focusing on touch gestures, on the one hand, children obtain average Q values over 65\% in gestures such as tap (Test 1) or drag-and-drop (Test 2), which also means a higher motor and cognitive capacity to perform the tests related to touch gestures. On the other hand, gestures such as pinch (Test 3 and 4) still remain a challenge at this age, with average Q values below 38\%. 

Finally, focusing on children aged 5 to 8 years old (Groups 6 to 8), we can observe similar trends in most tests of ChildCIdbLong. The average Q values for the tests where the pen stylus is used (Tests 5 and 6) are closer to the ones of adults (80.6\% vs. 94.7\% and 91.2\% vs. 95.6\%, respectively). This indicates that at this age children have similar motor and cognitive skills as adults to perform stylus gestures such as following a line and drawing the tree without going out of the outline. However, this trend is not observed for the case of tests involving touch gestures (Tests 1 to 4). In Tests 1 and 2, children achieve an average Q value of around 70\%, compared to almost 90\% for adults. We hypothesise this might be produced not by the motor factor, as children have the necessary motor skills to perform tap and drag-and-drop gestures at this age, but by the cognitive factor involving aspects such as reasoning and reaction time. Regarding Tests 3 and 4, we can see that the average Q value for children is around 45\%, much better than previous age groups (i.e., around 30\%). However, adults still obtain an average Q value of around 70\%. These results indicate that even for the control group it is difficult to achieve a result close to 100\% (i.e., ideal case). This may be produced due to the design of the tests as: \textit{i)} the pinch gesture requires greater fine motor skills, high finger coordination, and precise perception of force, and \textit{ii)} it is difficult to fit the rabbit between the red circles using a single tap and in a short period of time, even for adults with fully motor and cognitive developed skills.

\subsection{Longitudinal Analysis}

As we commented in Sec.~\ref{subsec:longitudinal}, nowadays there is a lack of longitudinal studies that automatically analyse the correct children's development as they grow up, using quantitative information captured through the interaction with mobile devices. In order to shed some light on this aspect, in this section we analyse how the proposed Q metric and the tests included in ChildCIdbLong can be used to measure the motor and cognitive skills of the children. For a better understanding of the results, we propose to use percentile-based growth representations, as indicated in Sec.~\ref{sec:methods}. This will allow an interesting comparison of the Q value for a specific child with the general population at their age.

Fig.~\ref{fig:longitudinal} shows an example of the evolution of the Q values achieved in Test 6 for two different children: \textit{i)} a TD child without apparent physical/cognitive impairments (blue diamonds), and \textit{ii)} a NTD child with special educational needs and an expressive language disorder (red circles). In the middle of the figure, we can see the percentile-based growth representation associated with Test 6 with its corresponding 10th, 50th, and 90th percentile limits. The analysis of both children is carried out for the 6 longitudinal acquisitions included in ChildCIdbLong. In particular, both children had their first interaction with the tests at 2-3 years old (Group 3) in the academic year 2019/20 and their last interaction at 5-6 years old (Group 6) in the academic year 2022/23, 3 years later. In the proposed representation we can observe how the Q value of the children evolves over time, extracting very interesting conclusions related to the motor and cognitive skills of the children. For completeness, at the top and bottom of Fig.~\ref{fig:longitudinal} we can also see from left to right the 6 graphical executions of the tree, each corresponding with one acquisition in time. Each representation has its associated Q value (\% visible in the lower right corner) along with an identifier number to position it on the growth chart. Due to the lack of space, similar representations will be included in the corresponding GitHub repository\dbfootnotelong.

On the one hand, focusing on the evolution of the TD child (blue diamonds) we can observe that the Q value has an increasing trend, being at all times above the 50th percentile and even above the 90th percentile when he was 3-5 years old (Group 4). This indicates, as confirmed by the specialists at the school, that their hand-eye coordination, fine motor skills, planning and organisation for colouring seem to be developing well. On the other hand, regarding the evolution of the NTD child (red circles), we can observe that most of the time the Q value is below the 50th percentile. This may be normal, as each child develops faster in some areas than in others. However, in the 5th data acquisition (Group 5), the Q value is below the 10th percentile, which could indicate that some aspects of their motor and cognitive skills are not developed correctly, as confirmed by the specialist of the school. In our opinion, the proposed tests included in ChildCIdbLong, together with the proposed Q metric, could be used as an automatic and quantitative tool for paediatricians, therapists, and specialists in schools to allow early detection of potential impairments, enabling early action and improving the quality of the rest of the child's life.

 
\section{Conclusion and Future Work}\label{sec:conclusion}

This study proposed a novel automatic quantitative metric called Test Quality (Q) designed to measure the motor and cognitive development of children through their interaction with a tablet device. In particular, children aged between 18 months and 8 years old interact with the 6 different tests presented in our ChildCIdbLong database, the largest publicly available longitudinal database to date for research in the Children-Computer Interaction (CCI) area. Each test requires different touch/stylus gestures and different motor and cognitive skills to be completed correctly. Therefore the Q metric quantitatively measures different abilities such as praxical skills related to visual recognition aspects and frontal executive patterns, among others.

Along with the Q metric, percentile-based growth representations are defined for each test, allowing to compare children in a two-dimensional space according to their development with respect to the typical age skills of the population (similar to the standards growth values for height, length, weight, and body mass index-for-age based on worldwide data of healthy children). The results obtained show the potential of the longitudinal database ChildCIdbLong in conjunction with the Q metric to measure whether the motor and cognitive development of children is correct or not over time. In addition, this metric could be useful for child experts (e.g., paediatricians, educators, psychologists, or neurologists) as a support tool to detect quickly and easily possible physical/cognitive impairments in children's development.

Future work will be oriented to: \textit{i)} identify whether there is any relationship between children's motor and cognitive development, their interaction with mobile devices and children's metadata (grades, prematurity, etc.), \textit{ii)} consider the ChildCIdbLong in other research areas related to e-Health and e-Learning, \textit{iii)} expand ChildCIdbLong with more participants and acquisition data, and \textit{iv)} define new variables associated with classical cognitive domains in order to assess children's attentional, visuo-spatial, and executive skills.

\newpage
\section*{Acknowledgements}

This work has been supported by project INTER-ACTION (PID2021- 126521OB-I00 MICINN/FEDER) and HumanCAIC (TED2021-131787BI00 MICINN). This is an on-going project carried out with the collaboration of the school GSD Las Suertes in Madrid, Spain.

{
\bibliographystyle{IEEEtran}
\bibliography{refs}
}

\vspace{-12mm}

\begin{IEEEbiography}[{\includegraphics[width=1in,height=1.25in,clip,keepaspectratio]{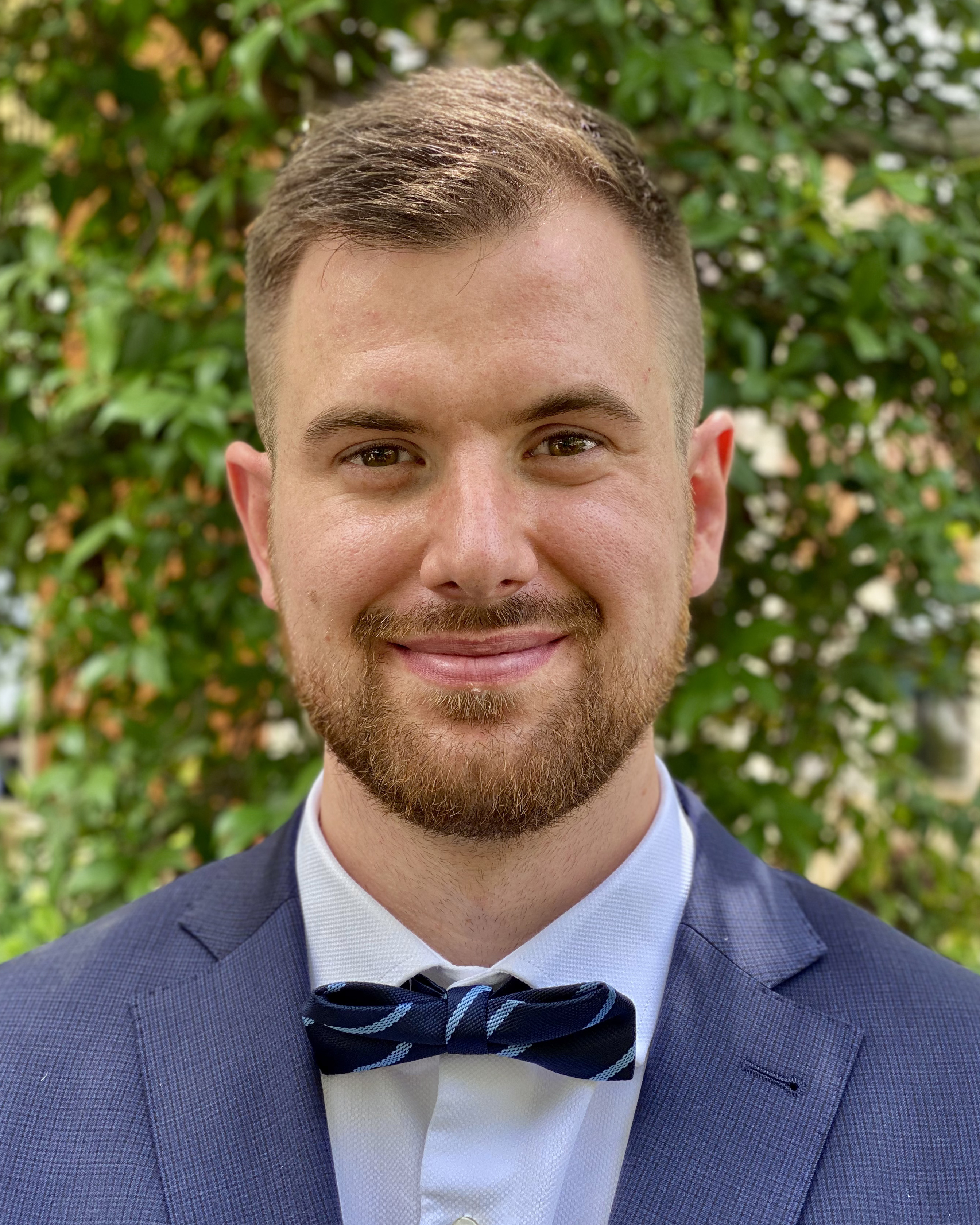}}]{Juan Carlos Ruiz-Garcia} received his B.Sc. degree in Computer Science Engineering in 2019 from the Universidad de Granada and got the M.Sc. degree in Research and Innovation in 2021 with the award of excellence from the Universidad Autonoma de Madrid, where he is currently pursuing a PhD degree in Computer and Telecommunication Engineering. In addition, in April 2020, he joined the Biometrics and Data Pattern Analytics - BiDA Lab as Pre-Doctoral Researcher at the same university. His research interests are mainly focused on the use of machine learning for e-Learning, e-Health, Human-Computer Interaction (HCI), and automatic Fall Detection Systems (FDS).
\end{IEEEbiography}

\vspace{-12mm}

\begin{IEEEbiography}[{\includegraphics[width=1in,height=1.25in,clip,keepaspectratio]{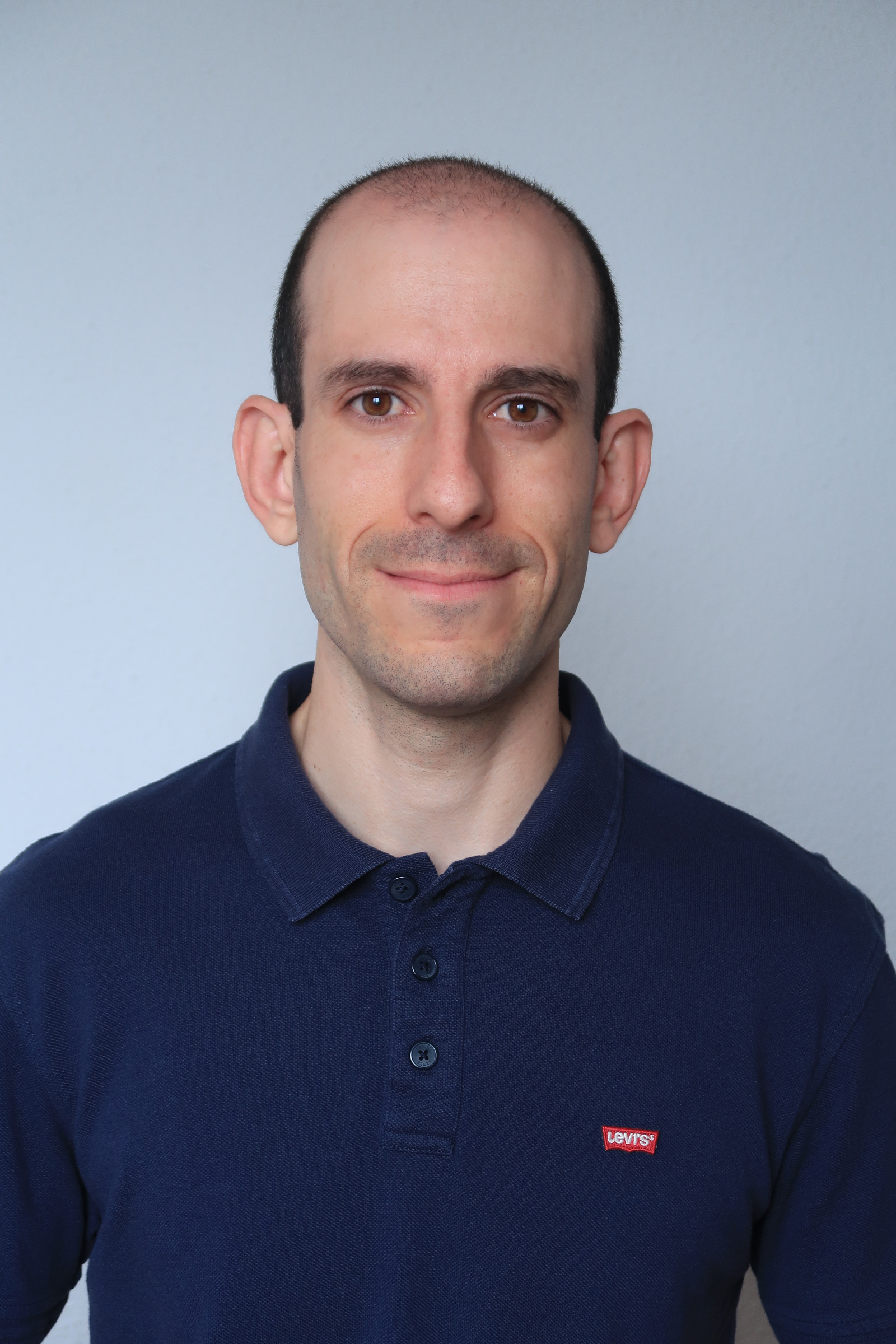}}]%
{Ruben Tolosana}
received the M.Sc. degree in Telecommunication Engineering, and his Ph.D. degree in Computer and Telecommunication Engineering, from Universidad Autonoma de Madrid, in 2014 and 2019, respectively. In 2014, he joined the Biometrics and Data Pattern Analytics - BiDA Lab at the Universidad Autonoma de Madrid, where he is currently collaborating as a PostDoctoral researcher. Since then, Ruben has been granted with several awards such as the FPU research fellowship from Spanish MECD (2015), and the European Biometrics Industry Award (2018). His research interests are mainly focused on signal and image processing, pattern recognition, and machine learning, particularly in the areas of DeepFakes, HCI, and Biometrics. He is author of several publications and also collaborates as a reviewer in high-impact conferences (WACV, ICPR, ICDAR, IJCB, etc.) and journals (IEEE TPAMI, TCYB, TIFS, TIP, ACM CSUR, etc.).
\end{IEEEbiography}

\vspace{-12mm}

\begin{IEEEbiography}[{\includegraphics[width=1in,height=1.25in,clip,keepaspectratio]{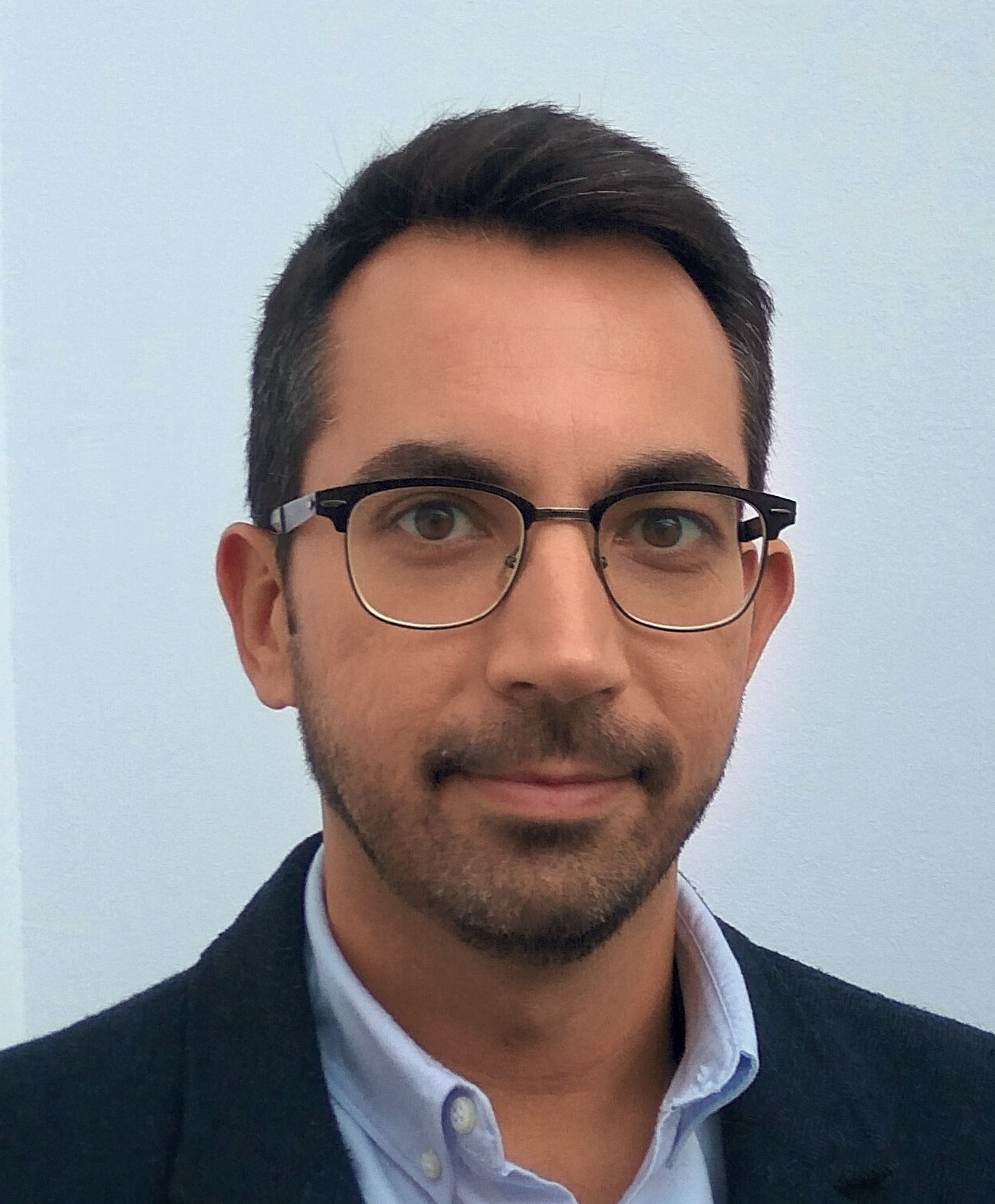}}]{Ruben Vera-Rodriguez} received the M.Sc. degree in telecommunications engineering from Universidad de Sevilla, Spain, in 2006, and the Ph.D. degree in electrical and electronic engineering from Swansea University, U.K., in 2010. Since 2010, he has been affiliated with the Biometric Recognition Group, Universidad Autonoma de Madrid, Spain, where he is currently an Associate Professor since 2018. His research interests include signal and image processing, pattern recognition, HCI, and biometrics, with emphasis on signature, face, gait verification and forensic applications of biometrics. Ruben has published over 100 scientific articles published in international journals and conferences. He is actively involved in several National and European projects focused on biometrics. Ruben has been Program Chair for the IEEE 51st International Carnahan Conference on Security and Technology (ICCST) in 2017; and the 23rd Iberoamerican Congress on Pattern Recognition (CIARP 2018) in 2018.
\end{IEEEbiography}

\begin{IEEEbiography}[{\includegraphics[width=1in,height=1.25in,clip,keepaspectratio]{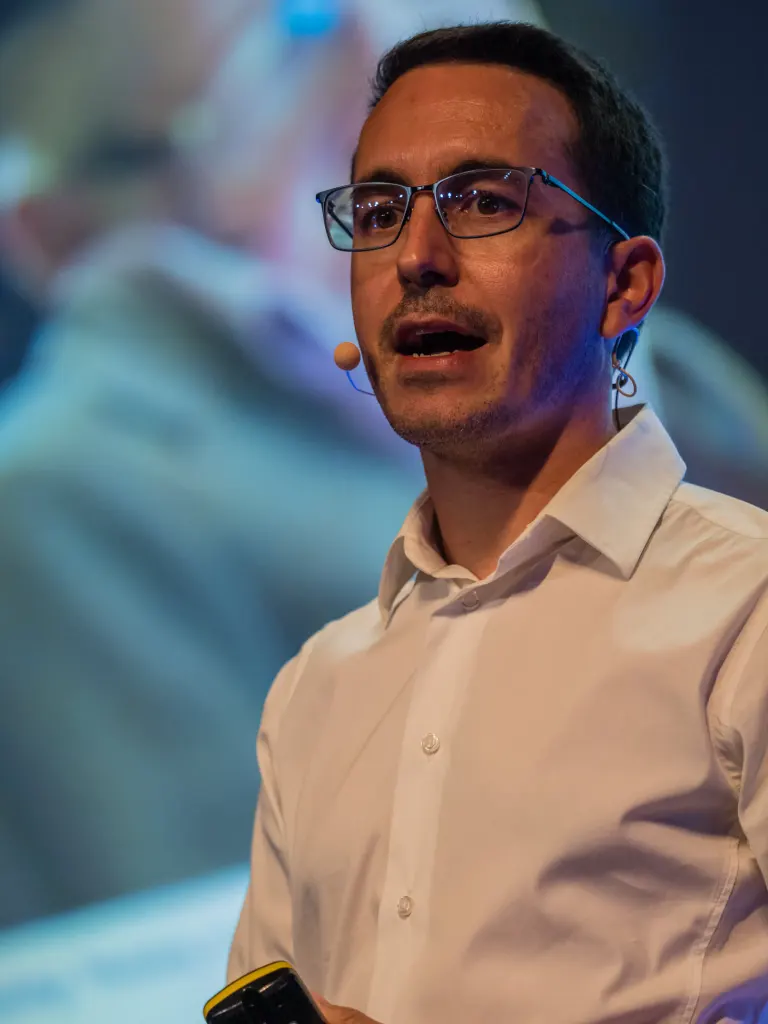}}]{Aythami Morales} received his M.Sc. degree in Telecommunication Engineering in 2006 from ULPGC. He received his Ph.D degree from ULPGC in 2011. He performs his research works in the BiDA Lab at Universidad Autónoma de Madrid, where he is currently an Associate Professor. He has performed research stays at the Biometric Research Laboratory at Michigan State University, the Biometric Research Center at Hong Kong Polytechnic University, the Biometric System Laboratory at University of Bologna and Schepens Eye Research Institute. His research interests include pattern recognition, machine learning, trustworthy AI, and biometrics. He is author of more than 100 scientific articles published in international journals and conferences, and 4 patents. He has received awards from ULPGC, La Caja de Canarias, SPEGC, and COIT. He has participated in several National and European projects in collaboration with other universities and private entities such as ULPGC, UPM, EUPMt, Accenture, Unión Fenosa, Soluziona, and BBVA.
\end{IEEEbiography}

\vspace{-20mm}

\begin{IEEEbiography}[{\includegraphics[width=1in,height=1.25in,clip,keepaspectratio]{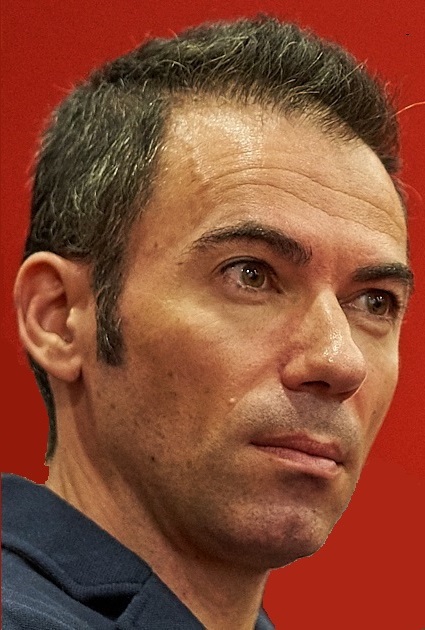}}]{Julian Fierrez} received the MSc and the PhD degrees from Universidad Politecnica de Madrid, Spain, in 2001 and 2006, respectively. Since 2004 he is at Universidad Autonoma de Madrid, where he is Associate Professor since 2010. His research is on signal and image processing, AI fundamentals and applications, HCI, forensics, and biometrics for security and human behavior analysis. He is Associate Editor for Information Fusion, IEEE Trans. on Information Forensics and Security, and IEEE Trans. on Image Processing. He has received best papers awards at AVBPA, ICB, IJCB, ICPR, ICPRS, and Pattern Recognition Letters; and several research distinctions, including: EBF European Biometric Industry Award 2006, EURASIP Best PhD Award 2012, Miguel Catalan Award to the Best Researcher under 40 in the Community of Madrid in the general area of Science and Technology, and the IAPR Young Biometrics Investigator Award 2017. Since 2020 he is member of the ELLIS Society.
\end{IEEEbiography}

\vspace{-20mm}

\begin{IEEEbiography}[{\includegraphics[width=1in,height=1.25in,clip,keepaspectratio]{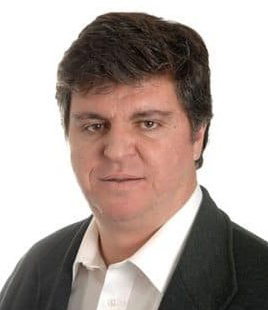}}]{Javier Ortega-Garcia} (Fellow, IEEE) received the M.Sc. degree in electrical engineering and the Ph.D. degree (cum laude) in electrical engineering from Universidad Politecnica de Madrid, Spain, in 1989 and 1996, respectively. He is currently a Full Professor at the Signal Processing Chair in Universidad Autonoma de Madrid—Spain, where he holds courses on biometric recognition and digital signal processing. He is a founder and Director of the BiDA-Lab, Biometrics and Data Pattern Analytics Group. He has authored over 300 international contributions, including book chapters, refereed journal, and conference papers. His research interests are focused on biometric pattern recognition (on-line signature verification, speaker recognition, human-device interaction) for security, e-Health and user profiling applications. He chaired Odyssey-04, the Speaker Recognition Workshop, ICB-2013, the 6th IAPR International Conference on Biometrics, and ICCST2017, the 51st IEEE International Carnahan Conference on Security Technology.
\end{IEEEbiography}

\vspace{-20mm}

\begin{IEEEbiography}[{\includegraphics[width=1in,height=1.25in,clip,keepaspectratio]{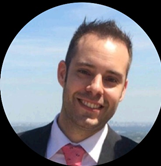}}]{Jaime Herreros-Rodriguez (JHR)} received the degree in Medicine in 2006 from Universidad Autónoma de Madrid, the tittle of neurologist in 2010 and he was awarded the title of Doctor in Medicine from the Universidad Complutense de Madrid (2019) with a distinction Cum Laude given unanimously for his doctoral thesis on migraine. He is also author of several publications in migraine and parkinsonism. He has collaborated with different research projects related to many neurological disorders, mainly Alzheimer and Parkinson's disease. JHR is a neurology and neurosurgery proffesor in CTO group, since 2008.
\end{IEEEbiography}

\end{document}